\documentclass[prc,twocolumn,showpacs,nofootinbib,aps,longbibliography,superscriptaddress]{revtex4-1}
\usepackage[utf8]{inputenc}
\usepackage[T1]{fontenc}
\usepackage{color}
\usepackage{amsmath}
\usepackage{url}
\usepackage{graphicx}
\usepackage{amssymb}
\usepackage{appendix}
\usepackage[section]{placeins}
\usepackage{multirow}
\usepackage{dcolumn}
\usepackage{hyperref}
\usepackage{mathrsfs} 
\usepackage{bm}
\usepackage{xcolor}
\usepackage{soul}
\usepackage{array}
\usepackage{lipsum}
\usepackage{relsize}
\usepackage{amsfonts}
\usepackage{comment}
\usepackage{outlines}
\usepackage[normalem]{ulem}
\usepackage{natbib}
\usepackage{booktabs}

\definecolor{linkcolor}{rgb}{0,0,0.40} 
\hypersetup{%
    pdfsubject=Paper,
    pdfkeywords={nuclear physics} {Bayesian} {chiral EFT},
    unicode = true,
    breaklinks = true,
    colorlinks = true,
    linkcolor = linkcolor,
    citecolor = linkcolor,
    menucolor = linkcolor,
    urlcolor = linkcolor
}

\newcommand{\obs}{\boldsymbol{y}}
\newcommand{\emumean}{\boldsymbol{\mu}}
\newcommand{\emucov}{\boldsymbol{\Sigma}_\mathrm{Emu}}
\newcommand{\params}{\boldsymbol{\omega}}
\newcommand{\singleparam}{\omega}
\newcommand{\vars}{\boldsymbol{\alpha}}

\newcommand{\optpot}{\mathrm{OP}}

\newcommand{\numparams}{10}
\newcommand{\NSnapshots}{N_\phi}
\newcommand{\NPotentials}{N_U}

\newcommand{\phihat}{\widehat\phi}

\newcommand{\PCAPOD}{\text{PCA}}

\newcommand{\trueparams}{\boldsymbol{\omega^t}}
\newcommand{\priorparams}{\boldsymbol{\omega^\mathrm{pr}}}
\newcommand{\singleprior}{\singleparam^\mathrm{pr}}

\newcommand{\smatch}{s_\text{match}}

\begin{document}

\title{ROSE: A reduced-order scattering emulator for optical models}

\author{D. Odell}
% \altaffiliation{Odell and Giuliani are co-authors contributing equally to this work}
\email{dodell4@alum.utk.edu}
\affiliation{Department of Physics and Astronomy, Institute of Nuclear and Particle Physics, Ohio University, Athens, Ohio 45701, USA}

\author{P. Giuliani}

\email{giulianp@frib.msu.edu}
\affiliation{Facility for Rare Isotope Beams, Michigan State University, East Lansing, Michigan 48824, USA}
\affiliation{Department of Statistics and Probability, Michigan State University, East Lansing, Michigan 48824, USA}

\author{K. Beyer}
\email{beykyle@umich.edu}
\affiliation{Department of Nuclear Engineering and Radiological Sciences, University of Michigan, 
Ann Arbor, Michigan 48109, USA}

\author{M. Catacora-Rios}
\affiliation{Facility for Rare Isotope Beams, Michigan State University, East Lansing, Michigan 48824, USA}
\affiliation{Department of Physics and Astronomy, Michigan State University, East Lansing, Michigan 48824, USA}

\author{M. Y.-H. Chan}
\email{moses.chan@northwestern.edu}
\affiliation{Department of Industrial Engineering and Management Sciences, Northwestern
University, Evanston, Illinois 60208, USA}

\author{E. Bonilla}
\email{edgard@stanford.edu}
\affiliation{Department of Physics, Stanford University, Stanford, California 94305, USA}

\author{R. J. Furnstahl}
\email{furnstahl.1@osu.edu}
\affiliation{Department of Physics, The Ohio State University, Columbus, Ohio 43210, USA}

\author{K. Godbey}
\email{godbey@frib.msu.edu}
\affiliation{Facility for Rare Isotope Beams, Michigan State University, East Lansing, Michigan 48824, USA}

\author{F. M. Nunes}
\email{nunes@frib.msu.edu}
\affiliation{Facility for Rare Isotope Beams, Michigan State University, East Lansing, Michigan 48824, USA}
\affiliation{Department of Physics and Astronomy, Michigan State University, East Lansing, Michigan 48824, USA}

\date{\today}

\begin{abstract}
A new generation of phenomenological optical potentials requires robust calibration and uncertainty quantification, motivating the use of Bayesian statistical methods.
These Bayesian methods usually require calculating observables for thousands or even millions of parameter sets, making fast and accurate emulators highly desirable or even essential.
Emulating scattering across different energies or with interactions such as optical potentials is challenging because of the non-affine parameter dependence, meaning the parameters do not all factorize from individual operators.
Here we introduce and demonstrate the Reduced Order Scattering Emulator (ROSE) framework, a reduced basis emulator that can handle non-affine problems.
ROSE is fully extensible and works within the publicly available BAND Framework software suite for calibration, model mixing, and experimental design. 
As a demonstration problem, we use ROSE to calibrate a realistic nucleon-target scattering model through the calculation of elastic cross sections.
This problem shows the practical value of the ROSE framework for Bayesian uncertainty quantification with controlled trade-offs between emulator speed and accuracy as compared to high-fidelity solvers.
 Planned extensions of ROSE
are discussed. 
\end{abstract}

\maketitle

\section{Introduction}

From nuclear reactions we can learn basic information about nuclei, such as energy levels, the distribution of matter inside the nucleus, and electromagnetic properties. To unfold the desired information from reaction experiments we must use a validated reaction model.
Current ab initio methods for reactions, starting from a many-body Hamiltonian with a nuclear force derived from chiral effective field theory, are capable of quantitatively describing reactions with light ions (see as an example~\cite{hebborn2022}). However, these methods cannot be applied to most reactions of interest involving heavier systems. In those cases, the typical approach is to reduce the many-body method to a few-body problem, for which the dynamics can be solved exactly (e.g.,~\cite{potel2017,johnson2020}). A key ingredient in few-body reaction theory is the optical potential. Whether we are interested in transfer reactions to study exotic nuclei~\cite{kay2022}, knockout reactions to understand quenching effects~\cite{aumann2021}, or breakup reactions for astrophysics~\cite{wang2021}, this ingredient is ubiquitous.

While optical potentials are pervasive in the field of nuclear reactions, they are one of the greatest source of uncertainty in reaction analyses. A recent overview~\cite{hebborn2023optical} stresses the importance of optical potentials for reactions
% over the field
and calls for a new global potential valid across the whole nuclear chart, including for nuclei away from stability, and with quantified uncertainties.
Several studies have been performed on quantifying uncertainties for the nucleon optical potentials~\cite{lovell2018,king2019,Lovell_2020,catacora2019,catacora2021,pruitt2023uncertainty} within a Bayesian framework~\cite{phillips2021get}.
Such studies have shown that uncertainties associated with the optical potential are substantial and Bayesian analyses can help in designing experiments to reduce those uncertainties. 

Although the two-body scattering problem can be solved rapidly
using traditional methods, when performing Bayesian studies, which require hundreds of thousands or millions of evaluations, computations become time consuming. 
In such cases, it is advantageous or even essential to use emulators as alternative computation methods. 
There have been several new developments of emulators for scattering~\cite{furnstahl2020efficient,drischler2021toward,melendez2021fast,Zhang:2021jmi,melendez2022model} based on reduced order methods. All of these emulators work best if the dependence on the model parameters is affine; they become less efficient when non-linear parameters are involved, as is the case for optical potentials.

In this paper, we demonstrate a new emulator for two-body scattering based on the Galerkin formulation through the reduced basis method (see Refs.~\cite{bonilla2022training,Drischler:2022ipa, quarteroni2015reduced,hesthaven2016certified} and references therein). The reduced basis method, equipped with the empirical interpolation method~\cite{barrault2004empirical,grepl2007efficient}, dramatically improves the speed of the calculations even
when the problem is non-affine. Typical non-affine parameters are the radius and diffuseness of the Woods-Saxon (and related) form-factors, which are employed in modeling individual terms in many optical potentials, as well as the energy dependence of the scattering solutions in general.
We introduce the associated software, called ROSE \cite{ROSE2023}, that is an integrated part of the publicly available BAND software suite for calibration, model mixing, and experimental design \cite{bandframework}.
The performance of the emulator is explored and assessed by calibrating a realistic optical potential by constraining elastic cross sections. 

We organize the rest of the manuscript
as follows. In Sec.~\ref{sec:formalism} we review the formalism and high-fidelity solvers used for nuclear scattering with a phenomenological (local) optical potential, and summarize the procedure for Bayesian calibration using an emulator.
We detail the reduced basis method (RBM) as applied to this scattering problem, including the construction of the reduced basis, the Galerkin projection for creating the reduced equations, and the method used to handled non-affine parameter dependence.
In section~\ref{sec:results} we illustrate the speed and accuracy trade-offs of ROSE compared to high-fidelity solvers, we detail the Bayesian calibration setup for the demonstration problem, and we present the results and their interpretation.
Our conclusions and outlook are summarized in Sec.~\ref{sec:conclusion-outlook}. 
The appendices contain additional details about the nucleon-target scattering equations, the posterior sampling procedure and results, and the search for emulator anomalies (see \cite{drischler2021toward}).

\section{Formalism} \label{sec:formalism}

\subsection{Nuclear Scattering with an optical potential}

In this section we briefly outline the formalism relevant for two-body scattering of a projectile nucleon ($A=1$) with spin $1/2$ with a target nucleus with total spin 0. For more details see Appendix~\ref{app:scattering} and Refs.~\cite{Thompson2009nuclear,herman2007empire}. 

The radial part of the Schrodinger scattering equation with a local potential is:
\begin{equation}
\begin{aligned}\label{Eq: Scattering Hamiltonian}
\Big({-}\frac{d^2}{dr^2}+&\frac{\ell(\ell+1)}{r^2} +\frac{2\eta k}{r} \\
& \null +\frac{2\mu}{\hbar^2}
V(r;\params)
-k^2\Big)\phi(r)=0,
\end{aligned}
\end{equation}
where the system has a reduced mass $\mu$, $\ell$ represents the angular momentum quantum number, and $\hbar k $ is the asymptotic linear momentum related to the energy through $E=\hbar^2k^2/2\mu$. For a charged system, the Coulomb interaction potential is characterized by the Sommerfeld parameter $\eta= Z_1 Z_2e^2 \mu /\hbar^2 k$, with charges $Z_1$ and $Z_2$ for the projectile and target, respectively. The nuclear short-range potential $V(r;\params)$, with parameters $\params$, characterizes the effective nuclear interaction. The potential $V$ could also depend on the orbital angular momentum $\ell$ and total spin $j$ of the system through spin-orbit coupling terms. 

For the optical potential we focus on in this work, $\params$ is a list of ten parameters:
\begin{align} \label{eqn:params_definition}
    \params=\{V_v,W_v,R_v,a_v,W_d,R_d,a_d,V_{so},R_{so},a_{so}\},
\end{align}
that characterize the strength, radius, and diffuseness of real and imaginary volume, imaginary surface, and real spin-orbit Woods-Saxon terms:
\begin{align}\label{Eq: Optical Potential}
      V(r;\params,\ell,j) = & \, {-}(V_v + iW_v) f_\text{WS}(r,R_v,a_v) \notag\\
      &  -i4a_dW_d \frac{d}{dr}f_\text{WS}(r,R_d,a_d) \notag\\
      & +2\ell\cdot s V_{so}  \Big(\frac{\hbar}{m_\pi c}\Big)^2 \frac{1}{r} \frac{d}{dr}f_\text{WS}(R_{so},a_{so}).
\end{align}
\noindent
The Woods-Saxon function is defined as:
\begin{equation}
    f_\text{WS}(r,R,a) = \Big[1+\text{exp}\Big( \frac{r-R}{a}\Big)  \Big]^{-1}.
\end{equation}
For the spin-orbit part, since we are only considering spin-zero targets with spin-$1/2$ projectiles, the coupling can only take the values:
\begin{equation}
2\ell\cdot s = \begin{cases}
  \ell  & \text{if }j=\ell+\frac{1}{2} \\
  -(\ell+1) & \text{if }j=\ell-\frac{1}{2}
\end{cases}  .
\end{equation}	
The arbitrary scale factor for the spin-orbit term uses the mass of the pion $m_\pi$ such that $\Big(\frac{\hbar}{m_\pi c}\Big)^2 \approx 2 \text{ fm}^2$.

It is convenient to re-scale Eq.~\eqref{Eq: Scattering Hamiltonian} by the change of variables $s\equiv kr$. With this re-scaling we define an operator $F_{\vars}$ using the same notation as in \cite{bonilla2022training}:
\begin{equation}\label{Eq: Scaled Scattering Hamiltonian}
    \begin{aligned}
     F_{\vars}[\phi(s)]=
         \Big(&{-}\frac{d^2}{ds^2}+\frac{\ell(\ell+1)}{s^2}+\frac{2\eta}{s}\\ 
         & \null + U(s;\vars)-1\Big)
         \phi(s;\vars)=0,
    \end{aligned}
\end{equation}
where the re-scaled nuclear potential
\begin{equation}
    U(s;\vars)=V(s/k,\params)2\mu/\hbar^2k^2
\end{equation}
now effectively depends on the energy, and the subscript $\vars$ is used to compactly represent the quantities that we wish to emulate across. 
For neutrons we emulate across energies and the potential parameters:
\begin{equation}\label{eq: vars}
    \vars \equiv \{\params, E\},
\end{equation}
while for protons, emulation is more complicated due to the scaling of $\eta$ with the energy. Therefore, in this work we only emulate the solution for protons across the potential parameters.
For the rest of the manuscript we will refer to the re-scaled solution $\phi(s;\vars)$ when writing $\phi(s)$, and to the re-scaled potential $U(s;\vars)$ when writing $U(s)$.

Once the numerical solutions $\phi(s)$ are obtained, phaseshifts and elastic cross sections angular distributions $d\sigma/d\Omega$ are obtained as described in Appendix~\ref{app:scattering}. These computations represent the observables that can be directly compared with experimental data, and will be the fundamental pillar of the Bayesian calibration defined in Sec.~\ref{subsec:bayescalib}.

The solutions to Eq.~\eqref{Eq: Scaled Scattering Hamiltonian} can be numerically computed by various methods.
We refer to the conventional ways of computing the solution as  ``high-fidelity solvers'' throughout the rest of the manuscript.
These include:
\begin{itemize}
    \item Methods that integrate $d^2 \phi/ds^2 \,$ in a discretized coordinate basis ($r$ or $s$), imposing initial conditions at $s \rightarrow 0$; these include the Numerov \cite{10.1093/mnras/84.8.592} and Runge-Kutta \cite{kutta1901beitrag, runge1895numerische} methods.
    \item Calculable R-matrix methods, which expand $F_{\bm{\alpha}}$ in a convenient pre-selected basis of functions and impose asymptotic boundary conditions at the channel matching radius $s = \smatch$ (see Appendix~\ref{app:scattering}) \cite{lane1958r, descouvemont2010r}. For scattering problems, a basis of Lagrange-Legendre functions is typically employed due to their compact support \cite{baye2015lagrange}.
\end{itemize}

We use the implementation of Runge-Kutta in \texttt{scipy.integrate.solve\_ivp} to generate results in Sec.~\ref{sec:results}  \cite{2020SciPy-NMeth, dormand1980family}.

High-fidelity solvers will usually have control parameters that can be used to tune the precision of the solution obtained, providing a trade-off between accuracy and speed; in the case of the adaptive-step Runge-Kutta implementation in \texttt{scipy}, these are the relative and absolute error tolerances in $\phi(s)$ used to determine the step $\Delta s$.  The solver propagates an initial condition of the function and its derivative $\{\phi(s_0),\phi'(s_0)\}$ up to a maximum value $s_\text{max}$, such that the short range potential vanishes $U(s)\approx 0$ for $s\ge s_\text{max}$. The starting value $s_0>0$ is chosen such that $\phi(s)$ is well approximated by its power behavior (see Eq.~\eqref{Eq: small s}) for $s\leq s_0$.  
Unless otherwise specified, we use an absolute and relative tolerance of $10^{-9}$ in \texttt{solve\_ivp}.

\subsection{Bayesian calibration with an emulator} \label{subsec:bayescalib}

The calibration of the optical potential Eq.~\eqref{Eq: Optical Potential}, in the Bayesian setting, refers to the use of physical observations to obtain probability distributions for the input parameters $\params$.  Denoting the observations as $\obs$, the statistical model relating the observations and the optical potential ($\optpot$) is of the form 
\begin{equation}\label{eq: obs}
    \obs = \optpot(\params) + \boldsymbol{\varepsilon},
\end{equation}
where OP$(\params)$ represents any observable that could be calculated from the optical model (such as differential cross sections),  and $\boldsymbol{\varepsilon} \sim \mathrm{N}(\boldsymbol{0}, \boldsymbol{\Sigma})$ is the residual error that we assume follows a multivariate Gaussian distribution with mean $\boldsymbol{0}$ and covariance matrix $\boldsymbol{\Sigma}.$  

Let $p(\cdot)$ generically denote a probability density and let $p(x|y)$ represent the conditional probability density of $x$ given $y$.  The process of Bayesian calibration then refers to identifying the conditional probability density $p(\params|\obs)$, the \textit{posterior} of $\params$.  Given $p(\params)$, the \textit{prior} of $\params$, the posterior is found via applying Bayes' theorem, i.e., 
\begin{equation} \label{eqn:bayes-posterior}
    p(\params|\obs) = \frac{p(\obs|\params) p (\params)}{p(\obs)} \propto p(\obs|\params)p(\params),
\end{equation}
where $\propto$ denotes equality up to a multiplicative constant.  The conditional density $p(\obs|\params)$ is the \textit{likelihood} function that provides a measure of compatibility between $\optpot(\params)$, the output corresponding to parameters $\params$, and the observations $\obs$, the measured angular distributions. For our study we construct the likelihood as: 
\begin{equation} \label{eqn:bayes-likelihood}
    \begin{split}
    p(\obs|\params) &\propto \\
    |\boldsymbol{\Sigma}|^{-\frac{1}{2}} &\exp\left\{ -\frac{1}{2}(\obs - \optpot(\params))^\mathsf{T} \boldsymbol{\Sigma}^{-1} (\obs - \optpot(\params))\right\},
    \end{split}
\end{equation}
where $|\boldsymbol{\Sigma}|$ is the determinant of the matrix $\boldsymbol{\Sigma}$. 

Following the statistical framework detailed in Sec.~4 of \cite{giuliani2023bayes}, once the posterior is specified we can build the \emph{predictive posterior} distribution of new unobserved data $\obs^{\textrm{pred}}$, given the already observed data, by marginalizing over the model parameters $\params$:
\begin{align} \label{eqn:predictive-posterior}
    p(\obs^{\textrm{pred}}|\obs) &= \int p(\obs^{\textrm{pred}}|\params) p(\params|\obs) \mathrm{d}\params.
\end{align}
This predictive posterior folds together the uncertainty on our model parameters coming from the posterior $ p(\params|\obs)$ Eq.~\eqref{eqn:bayes-posterior}, with the expected intrinsic error of new gathered data for a given model parameter $p(\obs^{\textrm{pred}}|\params)$, which we model as following a similar structure as our likelihood formulation Eq.~\eqref{eqn:bayes-likelihood}.

To estimate any of the distributions we described, Markov chain Monte Carlo (MCMC) algorithms \cite{gelman2013bayesian}, e.g., the Metropolis-Hastings algorithm, are commonly used for posterior sampling from Eq.~\eqref{eqn:bayes-posterior}.  In each iteration of such an algorithm, the likelihood function in Eq.~\eqref{eqn:bayes-likelihood} is evaluated at a different value of $\params$.
The computational burden becomes significant when hundreds of thousands of posterior samples of $\params$ are required, motivating the use of emulators (also known as surrogate models). The goal is then to build an emulator that can provide accurate predictions for all these samples as an efficient alternative to repeatedly solving the scattering equations ~\eqref{Eq: Scaled Scattering Hamiltonian}. For our specific case in this work, our target will be to build an emulator able to perform more than one million evaluations per hour on commodity hardware, while maintaining an accuracy of less than 10$\%$ error.

Once the emulator is built, we can modify the posterior Eq.~\eqref{eqn:bayes-posterior} to take into account that we are obtaining approximate calculations. For any parameter sample $\params$, we summarize the trained emulator by its prediction mean $\emumean(\params)$ and covariance $\emucov(\params)$. This covariance matrix characterizes the (correlated or uncorrelated) error that the emulator is making,
resulting in the approximate posterior
\begin{align}
\label{Eq: posterior emu}
    p(\params|\obs) \propto & \ |\boldsymbol{\Sigma} + \emucov(\params)|^{-\frac{1}{2}} \notag\\   &\times\exp\Big\{ -\frac{1}{2}\big(\obs - \emumean(\params)\big)^\mathsf{T} \big(\boldsymbol{\Sigma} + \emucov(\params)\big)^{-1} \notag \\
    &\times\big(\obs - \emumean(\params)\big)\Big\} p(\params),
\end{align}
where the emulator error and the observation error are assumed to be independent, leading to the additive covariance $\boldsymbol{\Sigma} + \emucov(\params)$. Since quantifying the uncertainty emulators such as ROSE is non-trivial, in this study we limit $\emucov$ in Eq.~\eqref{Eq: posterior emu} to be independent of $\params$, and estimate it empirically across a set of test parameters, as described later in Sec.~\ref{subsec:calibrationsetup}.

\subsection{The Reduced Basis Method emulator}\label{Sec: RBM}

\begin{figure}[!h]
    \centering{\includegraphics[width=\columnwidth]{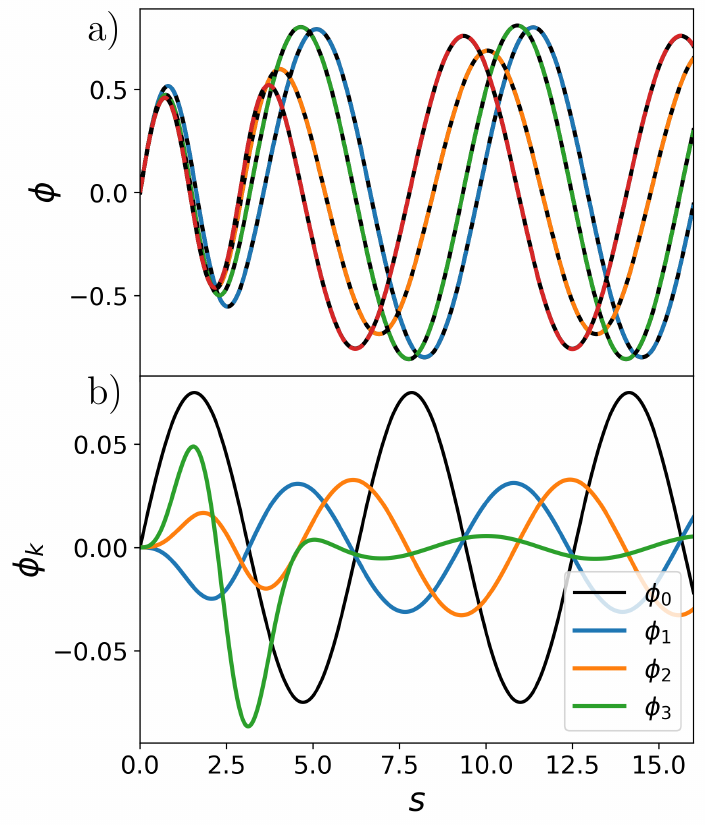}}
    \caption{Panel a): HF, $S$-wave solutions (solid, colored lines) alongside emulated solutions (dashed, black lines) at $E_{\rm c.m.} = $ 14 MeV $^{40}$Ca(n,n) and four different test parameter points. Panel b): Basis states for the expansion of $\phihat$ as defined in \eqref{Eq: RB}. The $\phi_0$ term (in black) represents the solution in the absence of an interaction $U(s;\vars)$, and it is used to enforce the boundary conditions at $s\sim 0$. The other three functions $\phi_k$ with $k=[1,2,3]$ (colored solid lines) are the first three principal components of the differences of snapshots with $\phi_0$ as defined in \eqref{Eq:POD_basis}. Only the real components are displayed.}
      
    \label{fig:PCA}
\end{figure}

Various emulator options are available for overcoming the computational burdens of Bayesian analysis, and each comes with different strategies and performances. Some emulators are less intrusive and require little knowledge of the underlying structure of the system of interest such as the scattering equation~\eqref{Eq: Scaled Scattering Hamiltonian}, for example. These emulators usually interpolate between inputs and outputs, for example the optical potential parameters Eq.~\eqref{eqn:params_definition} and the observables such as the differential cross section Eq.~\eqref{Eq: diff cs}.
Among such less intrusive emulators, the use of Gaussian processes has been commonly adopted since \textcite{kennedy2001bayesian}, see for examples \cite{mcdonnell2015uncertainty, wesolowski2019exploring, melendez2019quantifying, surer2022, chan2023surrogate, svensson2023inference},  because of both its straightforward implementation and its natural quantification of uncertainty.  In particular, \textcite{surer2022} have considered the emulation of cross-sections in nuclear breakup reactions using Gaussian processes and provided uncertainty quantification for four parameters.

On the other hand, more intrusive emulators can leverage physical information, e.g., the set of scattering equations, in their construction and thus retain structural knowledge of the system. Several such emulators have been developed within the field of model order reduction for wide applications in science and engineering in general~\cite{quarteroni2014reduced,brunton2016discovering, benner2017model,brunton2019data}, as well as in nuclear physics ~\cite{frame2018eigenvector,furnstahl2020efficient,melendez2021fast,drischler2021toward,konig2020eigenvector,Zhang:2021jmi,melendez2022model,garcia2023wave,sarkar2021convergence,sarkar2022self,demol2020improved,djarv2022bayesian,yapa2022volume,franzke2022excited,bonilla2022training,anderson2022applications,giuliani2023bayes,bai2021generalizing,ekstrom2019global} (see \cite{drnuclear,Drischler:2022ipa,FRIB2023Materials,Duguet:2023wuh} for pedagogical resources tailored to a nuclear physics audience). The reduced basis method~\cite{quarteroni2015reduced,hesthaven2016certified} has been a particularly successful framework, and ROSE, as we describe next, belongs to this category. 

If we restrict the scattering equation~\eqref{Eq: Scaled Scattering Hamiltonian} to a maximum value $s_\text{max}$, the formal exact solution $\phi$  exists in an infinite-dimensional Hilbert space\footnote{If we consider the entire interval $s\in [0,\infty)$ then the scattering solutions $\phi(s)$ do not have compact support and exist in a rigged Hilbert space instead.} $\mathcal{H}$, and the operator $F_{\vars}$ maps functions from $\mathcal{H}$ to itself. More precisely, $\phi$ exists within a manifold of $\mathcal{H}$ parametrized by all the variables $\vars$. When $\phi$ is represented as a list of values on a grid $\{\phi(s_1),\phi(s_2)...,\phi(s_\mathcal{N})\}$, it is still a vector within a possibly high dimensional space of size $\mathcal{N}$. 
Dimensionality reduction techniques like reduced order models usually work by first identifying and describing the system with a much smaller set of reduced coordinates $n_\phi\ll \mathcal{N}$, and then constructing governing equations for the new coordinates. 
The reduced basis method in particular does this by exploiting two linear subspaces of $\mathcal{H}$, one to restrict the input of $F_{\vars}$ (where $\phi$ resides), and the other to restrict its output~\cite{bonilla2022training}, usually through a Galerkin projection~\cite{rawitscher2018galerkin}\footnote{Alternatively, in formal mathematics this can be seen as a restriction of the weak formulation of the equations to a subspace of the Hilbert space~\cite{melendez2022model}.}. By working on these subspaces of smaller dimension $n_\phi$, a high computational efficiency can be obtained. We now proceed to describe how both steps, identifying reduced coordinates and constructing the reduce equations, are carried out within the ROSE framework. Chapters 2 and 3 of Ref.~\cite{drnuclear}, as well as the ROSE documentation~\cite{ROSE2023}, provide guided examples with interactive codes on these procedures for the specific case of Eq.~\eqref{Eq: Scaled Scattering Hamiltonian}.

\subsubsection{Training space: the reduced coordinates}

We find the first subspace by constructing an approximation to $\phi(s)$ through a reduced basis expansion:
\begin{equation}\label{Eq: RB}
    \begin{aligned}
    \phi(s;\vars) \approx \phihat(s;\vars)
     =\phi_0(s) + \sum_{k=1}^{n_\phi} a_k(\vars) \phi_k(s),     
    \end{aligned}
\end{equation}
where now the spatial dependence in $s$ is only carried out by the basis $\phi_k(s)$, while the coefficients $\boldsymbol{a}(\vars)=\{a_k\}$ can change to accommodate variations in the variables $\vars$. Each channel $(\ell,j)$ will have a different reduced basis, built using the same procedure.

The term $\phi_0(s)$ is an optional basis element without an assigned coefficient that helps enforce the initial conditions. As we discuss in Sec.~\ref{Sec: projecting}, such a term is crucial to create a non-homogeneous system of equations for the coefficients $\boldsymbol{a}$, and avoid obtaining the trivial solution $\phihat =0$. We select $\phi_0(s)$ as the Coulomb function $F_\ell(\eta,s)$ (see Appendix~\ref{app:scattering}), which is the solution to Eq.~\eqref{Eq: Scaled Scattering Hamiltonian} in the absence of a nuclear potential $U(s,\vars)$. We choose the rest of the basis $\phi_k(s)$ by performing a principal component analysis (PCA), which is related to the singular value decomposition algorithm~\cite{brunton2019data} and known in the reduced basis literature as 
proper orthogonal decomposition~\cite{quarteroni2015reduced}. We extract the principal components of the difference between the solution in the absence of the nuclear potential and $\NSnapshots$ ``snapshots'' $\phi(s;\params_m)$, which are high-fidelity solutions to Eq.~\eqref{Eq: Scaled Scattering Hamiltonian} for $\NSnapshots$ different values of the parameters $\params$ and possibly the energy:
\begin{equation}
     \{ \phi_k\}_{k=1}^{n_{\phi}} = \PCAPOD \Big[\{\phi(s,\vars_m)-\phi_0(s) \}_{m=1}^{\NSnapshots}\Big].\label{Eq:POD_basis}
\end{equation}

The proper orthogonal decomposition captures the $n_\phi$ most relevant directions of variability in the snapshots, and in this case it has the convenient interpretation of selecting the principal modes of variation -caused by the potential- around the ``free'' solution $\phi_0(s)$ .
Figure~\ref{fig:PCA} a) gives a visual display of the approximation's Eq.~\eqref{Eq:POD_basis} effectiveness in reproducing the HF solutions.
Panel b) contains the first three $\PCAPOD$ basis, as well as the $\phi_0$ term, upon which the accurate reproduction of the HF solutions are built.
From panel b), it is apparent that the variability is greatest at short distances, or small $s$, where the nuclear interaction contributes to the solution of the Schr\"odinger equation.

\subsubsection{Projecting space: the reduced equations}\label{Sec: projecting}

Once the reduced basis has been constructed, we need to create a system of equations to find the coefficients $\boldsymbol{a}$ as the parameters $\vars$ change. We achieve this by selecting the second subspace of $\mathcal{H}$ expanded by ``projecting'' (or ``test'') functions $\psi_j$ with $j\in [1,n_\phi]$. The output of the operator $F_{\vars}$ is restricted to this subspace, and satisfying Eq.~\eqref{Eq: Scaled Scattering Hamiltonian} (in this subspace) is done by requiring that the projection of the residual $F_{\vars}[\phihat]$ onto each of the $\psi_j$, for $j\in [1,n_\phi]$, is zero:
\begin{align}    \label{eqn:projections}
    \langle \psi_j|F_{\vars}[\phihat]\rangle &= \langle \psi_j | F_{\vars}|\phi_0 \rangle
    + \sum_{k=1} ^{n_\phi} a_k\langle \psi_j | F_{\vars}|\phi_k \rangle 
     \notag \\
    & =0
   .
\end{align}
Here we have used that the operator Eq.~\eqref{Eq: Scaled Scattering Hamiltonian} is linear: $F_{\vars}[\phihat] = F_{\vars}\phihat$. We use Dirac's notation for inner products: $\langle \psi|F_{\vars}| \phi \rangle = \int \psi^*(s) F_{\vars}\phi(s) ds$, where $\psi^*$ denotes the complex conjugate of $\psi$.
We select
\begin{equation}
    \psi_j(s)=\phi_j^*(s)
\end{equation}
for the projecting functions, a choice that connects the Galerkin-method approach with scattering emulators based on the Kohn-Variational-Principle \cite{furnstahl2020efficient,drischler2021toward}, as proven in \cite{bonilla2022training}. Note that with this choice, a double complex conjugate will cancel when plugging $\psi_j(s)=\phi_j^*(s)$ in equation \eqref{eqn:projections}. 

The $n_\phi$ equations~\eqref{eqn:projections} for the array of coefficients $\boldsymbol{a}$ can be written in matrix form: 
\begin{equation}\label{Eq: matrices}
   \boldsymbol{M}(\vars)\boldsymbol{a}=\boldsymbol{c}(\vars) ,
\end{equation}
where the $n_\phi{\times} n_\phi$ matrix $\boldsymbol{M}(\vars)$ is formed by the inner products between the projectors $\langle\psi_j|$ and the operator $F_{\vars}$ acting on each one of the basis $|\phi_k\rangle$, and $\boldsymbol{c}(\vars)$ is the non-homogenous array of size $n_\phi$ obtained from the projection of $F_{\vars}[\phi_0]$ onto the $\psi_j$:
\begin{equation}\label{Eq: integrals}
    \begin{aligned}
    M_{j,k} &= \langle\psi_j|F_{\vars}|\phi_k\rangle =\int \psi_j^*(s) F_{\vars} \phi_k(s) ds, \\ c_j&=-\langle\psi_j|F_{\vars}|\phi_0\rangle =-\int \psi_j^*(s) F_{\vars} \phi_0(s) ds.
    \end{aligned}
\end{equation}

The reduced basis $\phi_k$, and the system of equations for $\boldsymbol{a}$ involving $\boldsymbol{M}$ and $\boldsymbol{c}$, are usually computed only once in what is called the offline, or training, stage of the emulator. The online stage consists of then using the trained emulator to swiftly give an approximate solution for a new value of the parameters by solving the $n_\phi$-dimensional linear system in Eqs.~\eqref{Eq: matrices}~\cite{quarteroni2015reduced,hesthaven2016certified}. The rationale behind the Offline-Online division hinges on the fact that both $\phi(s)$ and $F_{\vars}$ reside in high-dimensional spaces of sizes $\mathcal{N}$ and  $\mathcal{N}^2$, respectively, while the approximation $\phihat(s)$ lives in a much smaller space of size $n_\phi\ll\mathcal{N}$. Computing each element of both $\boldsymbol{M}$ and $\boldsymbol{c}$ involves operations that scale with the grid size $\mathcal{N}$, but if they can be done only once, then the solution $\boldsymbol{a}=\boldsymbol{M}^{-1}\boldsymbol{c}$ involves only operations that scale with the reduced system size $n_\phi$ for every new values of the parameters. We describe how this is achieved for the non-affine parameters of the optical potential in the following discussion.

\subsubsection{The Empirical Interpolation Method}

Computing the integrals Eqs.~\eqref{Eq: integrals} for a general unspecified value of the parameters can be done if the operator $F_{\vars}$ is affine on those quantities. This is the case for the multiplicative strength parameter $V_v$ in Eq.~\eqref{Eq: Optical Potential}, since it factorizes from a function that depends on $r$ (or $s$), but it is not the case for the radius $R_v$ or the energy for the scaled potential $U(s;\vars)$ in Eq.~\eqref{Eq: Scaled Scattering Hamiltonian}.
To maintain a swift emulator that avoids operations that scale with $\mathcal{N}$, we can recover an affine dependence on $U(s;\vars)$ through the empirical interpolation method (EIM)~\cite{barrault2004empirical,grepl2007efficient,quarteroni2015reduced,hesthaven2016certified}. 

When Eq.~\eqref{Eq: Scaled Scattering Hamiltonian} is projected on a grid, $\phi(s)$ is represented as a vector of size $\mathcal{N}$ while the potential $U(s;\vars)$ is represented as a matrix of size $\mathcal{N}^2$ (diagonal, in the case of local optical potentials). In the same spirit as Eq.~\eqref{Eq: RB}, we seek a reduced dimensional representation of $U(s;\vars)$ by constructing an approximation through the sum of $n_U\ll\mathcal{N}^2$ terms:
\begin{equation}\label{Eq: potential RB}
    U(s,\vars) \approx \widehat U(s,\vars) = \sum _{i=1} ^{n_U} b_i(\vars) u_i(s).
\end{equation}
In the same way as the separation we did in Eq.~\eqref{Eq: RB}, the spatial dependence on $s$ is only carried out by the potential basis $u_i(s)$, while the coefficients $\boldsymbol{b}\equiv\{b_i(\vars)\}$ can change to accommodate variations in the variables $\vars$. 

The basis expansion $u_i(s)$ is computed once, following the same approach we did for constructing the reduced basis for $\phi$. We explicitly calculate $\NPotentials\ge n_U$ potentials $U(s;\vars)$ for $\NPotentials$ values of the parameters $\vars$,
perform a principal component analysis, and retain the $n_U$ most important components:
\begin{equation}\label{eq: potential PCA}
    \{ u_i(s)\}_{i=1}^{n_U} = \PCAPOD \Big[\{ U(s,\vars_m)\}_{m=1}^{\NPotentials}\Big].
\end{equation}

\begin{figure}[!t]
    \centering{\includegraphics[width=\columnwidth]{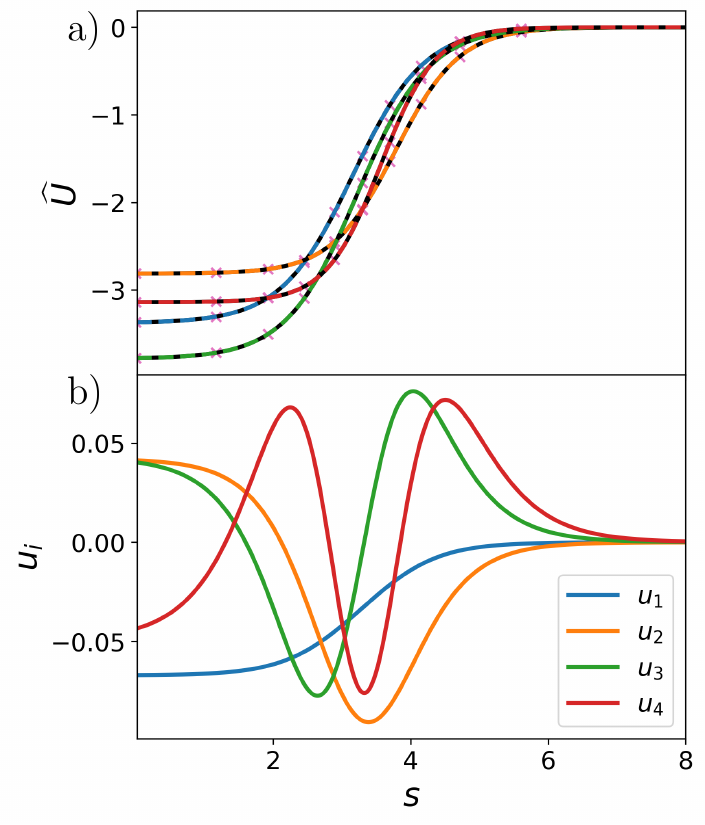}}
    \caption{Panel a): HF potentials (solid, colored lines) alongside EIM-emulated potential (dashed, black lines) at four different test parameter points. Pink crosses indicate the points $s_j$ at which agreement is enforced. Panel b): $\PCAPOD$ basis components of the emulated potential Eq.~\eqref{eq: potential PCA}. Only the real parts of the potential are shown. }      
    \label{fig:EIM}
\end{figure}

The reduced basis in Eq.~\eqref{eq: potential PCA} is computed only once, while the coefficients $\boldsymbol{b}(\vars)$ need to be determined for every new value of the parameters. This determination should be done involving operations that only scale with $n_U\ll\mathcal{N}^2$, to preserve efficiency. Within the EIM this can be done by selecting $n_U$ locations $s_j$ on which the approximation Eq.~\eqref{Eq: potential RB} is made exact:
\begin{equation}\label{Eq: EIM Eqs}
    U(s_j,\vars) -\sum _{i=1} ^{n_U} b_i(\vars) u_i(s_j) = 0, \quad \text{for }j\in [1,n_U].
\end{equation}
We can interpret that $\widehat U(s,\vars)$ is interpolating $U(s,\vars)$ through $s_j$ for all the other values of $s$. These equations for the coefficients $\boldsymbol{b}$ can also be interpreted as a Galerkin projection \eqref{eqn:projections} of the residual $ U(s,\vars)- \widehat U(s,\vars)$ on Dirac delta functions $\psi_j(s) = \delta(s-s_j)$ \cite{wang2022temporal,chen2021eim}. 

\begin{figure*}[!ht]
    \centering{\includegraphics[width=1\textwidth]{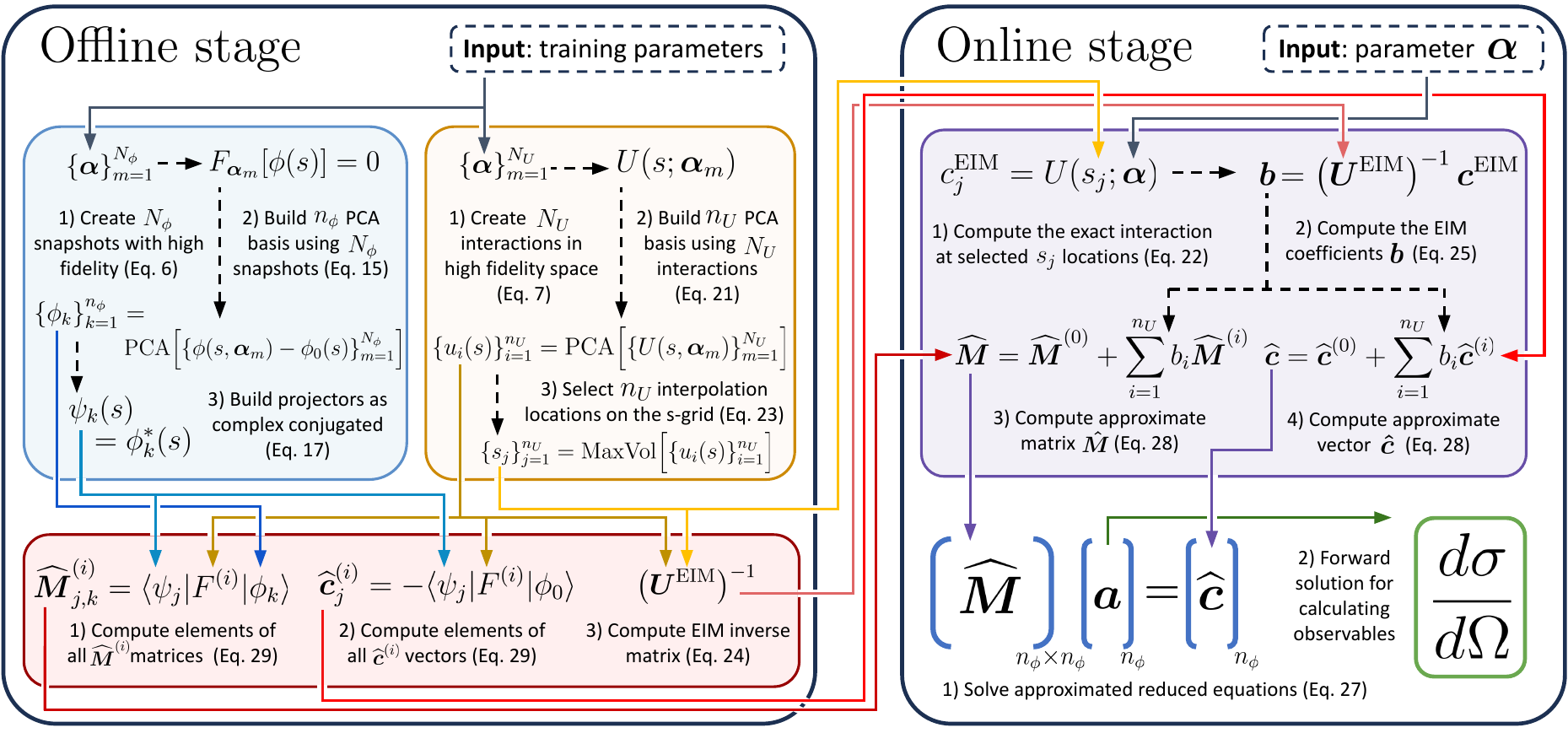}}
\caption{Flowchart illustrating the offline-online separation in ROSE. In the offline stage, for a fixed channel $(\ell,j)$, we create reduced bases by using Principal Component Analysis over the wavefunctions and potentials associated with sets of training parameters $\vars$. These reduced bases are used to create the building blocks of the approximate equations: the reduced operators $\boldsymbol{\widehat{M}}^{(i)}$, the reduced boundary terms $\boldsymbol{\widehat{c}}^{(i)}$, a set of interpolation points $s_j$, and the inverse of the empirical interpolation matrix $\boldsymbol{U}^{\text{EIM}}$. In the online stage, the emulator avoids performing operations that scale with the high-fidelity dimension $\mathcal{N}$: the query parameters $\vars$ are used to interpolate the potential at the selected locations $s_j$ to assemble the approximate reduced equations $\widehat{\boldsymbol{M}}\boldsymbol{a}=\widehat{\boldsymbol{c}}$. Solving this linear system for each channel $(\ell,j)$ provides the coefficients $\boldsymbol{a}(\vars)$ to be used for calculating observables for different query parameters.}    
    \label{fig:Flowchart}
\end{figure*}

To select the locations $s_j$ on which to do the interpolation, we follow the MaxVol algorithm \cite{Goreinov_2010}:
\begin{equation}
    \{ s_j\}_{j=1}^{n_U} = \text{MaxVol} \Big[ \{ u_i(s)\}_{i=1}^{n_U}  \Big]
\end{equation}
or DETMAX algorithm and its variants in experimental design \cite{mitchell2000algorithm, nguyen1992review, miller1994algorithm}, to obtain a `D-optimal' design.  Finding a D-optimal design is equivalent to minimizing the volume of the confidence ellipsoid for estimators of $\boldsymbol{b}$ \cite{nguyen1992review}.  Finding the set of global optimum locations is NP-hard \cite{welch1982algorithmic}, yet a local optimum can be efficiently identified using the aforementioned algorithms.  In practice, a local optimum is sufficient to recover 
a desirable interpolation accuracy.  We construct a matrix of size $\mathcal{N}\times n_U$ in which the columns are the $n_U$ basis $u(s)$ and the rows are the $\mathcal{N}$ grid points---and possible candidates $s_j$---in the $s$ variable. The algorithm seeks to choose the $n_U$ locations $s_j$ that maximize the determinant of the following reduced $n_U\times n_U$ matrix:
\begin{equation}
    \boldsymbol{U}^\text{EIM}=\begin{bmatrix}
        u_1(s_1) & u_2(s_1)  & ...  & u_{n_U}(s_1)  \\
        u_1(s_2)  & u_2(s_2)  & ...  & u_{n_U}(s_2)  \\
        \vdots  & \vdots  & \vdots  & \vdots  \\
        u_1(s_{n_U})  & u_2(s_{n_U})  & ...  & u_{n_U}(s_{n_U})  \\
    \end{bmatrix}_{n_U\times n_U}
\end{equation}

The algorithm is started by first selecting $n_U$ locations $s_j$ at random over the grid and then iteratively swapping them for other locations in a greedy fashion by comparing their expansion coefficients \cite{Goreinov_2010}.  

The results of the EIM are shown in Fig.~\ref{fig:EIM}.
Panel a) displays four examples of the exact and approximated potential curves where the visual overlap indicates that the method is accurate.
The interpolation points, $s_j$, are given as pink crosses. Panel b) shows the first four reduced basis $u_i(s)$ formed from the principal component analysis. Note the horizontal scale extends only to $s=\ 8$, since the interpolating points selected by MaxVol focus on areas where the value of the potential is most sensitive to changes in the parameter set.

Once the locations $s_j$ have been chosen, for a given $\bm{\alpha}$ the $n_U \times n_U$ linear system Eq.~\eqref{Eq: EIM Eqs} can be solved for the coefficients $\boldsymbol{b}(\vars)$. In practice, the inverse of the $n_U \times n_U$ matrix formed by $\boldsymbol{U}^\text{EIM}_{ji} \equiv u_i(s_j)$ is pre-computed in the offline stage since it is independent of $\vars$. In the online stage the exact potential can be evaluated at the interpolation points $c_j^\text{EIM}(\bm{\alpha}) = U(s_j, \bm{\alpha})$, and the coefficients determined by a simple matrix-vector multiplication:
\begin{equation}\label{eq: EIM coeffs}
    \bm{b}(\bm{\alpha}) = \left(\bm{U}^\text{EIM}\right)^{-1} \bm{c}^\text{EIM}(\bm{\alpha}).
\end{equation}

This determines a realization of the approximate affine decomposition in Eq.~\eqref{Eq: potential RB} for a given $\bm{\alpha}$. An approximate operator $\widehat{F}_{\vars}$ can then be constructed by substituting the approximation Eq.~\eqref{Eq: potential RB} in Eq.~\eqref{Eq: Scaled Scattering Hamiltonian}:
\begin{equation}\label{eq: approx operator}
    F_{\vars} \approx \widehat{F}_{\vars} = F^{(0)} + \sum_{i=1}^{n_U} b_i(\vars) F^{(i)},
\end{equation}
where $F^{(0)}$ represents the part of the original operator that is independent of the parameters, and $F^{(i)}=u_i(s)$ are the $n_U$ identified principal components of variations in the part of the operator that depends on the parameters Eq.~\eqref{eq: potential PCA}. 
An approximate version of Eq.~\eqref{Eq: matrices} can then be constructed by the projections $\langle \psi_j|\widehat{F}_{\vars}[{\phihat}]\rangle$ :
\begin{equation}\label{Eq: matrices hat}
   \widehat{\boldsymbol{M}}(\boldsymbol{b})\boldsymbol{a}=\widehat{\boldsymbol{c}}(\boldsymbol{b}),
\end{equation}
where now the approximate matrix $\widehat{\boldsymbol{M}}$ and approximate vector $\widehat{\boldsymbol{c}}$ consist of the sum of the projections of the $n_U+1$ operators in Eq.~\eqref{eq: approx operator} with the test functions $\psi_j$ on the reduced basis $\phi_k$, and with the $\phi_0$ term, respectively:
\begin{equation}\label{eq: matrix building}
    \begin{aligned}
    &\widehat{\boldsymbol{M}}(\boldsymbol{b}) = \widehat{\boldsymbol{M}}^{(0)} + \sum_{i=1}^{n_U} b_i(\vars)\widehat{\boldsymbol{M}}^{(i)}, \\
    &\widehat{\boldsymbol{c}}(\boldsymbol{b}) = \widehat{\boldsymbol{c}}^{(0)} + \sum_{i=1}^{n_U} b_i(\vars)\widehat{\boldsymbol{c}}^{(i)}.
    \end{aligned}
\end{equation}
Both quantities are now affine in the coefficients $\boldsymbol{b}$, and are constructed as:
\begin{equation}\label{Eq: integrals_EIM}
    \begin{aligned}
    \widehat{\boldsymbol{M}}_{j,k}^{(i)} &= \langle\psi_j|F^{(i)}|\phi_k\rangle =\int \psi_j^*(s) F^{(i)} \phi_k(s) ds, \\ 
    \widehat{\boldsymbol{c}}_j^{(i)}&=-\langle\psi_j|F^{(i)}|\phi_0\rangle =-\int \psi_j^*(s) F^{(i)} \phi_0(s) ds.
    \end{aligned}
\end{equation}
for $i\in [0,n_u]$. All such projections are calculated in the offline stage of the emulator.

In the online stage, when the emulator is deployed and evaluated for new parameters $\vars$, the coefficients in Eq.~\eqref{eq: EIM coeffs} are computed and the approximate matrix and vector are built by summing the pre-computed matrix terms in Eqs.~\eqref{eq: matrix building}. Finally, the approximate system of equations~\eqref{Eq: matrices hat} are solved for the coefficients $\boldsymbol{a}$ of the reduced basis expansion Eq.~\eqref{Eq: RB} and then used for computing observables. None of these operations on the online stage scale with the original high dimension size $\mathcal{N}$. Figure~\ref{fig:Flowchart} presents the ROSE flowchart for the offline-online separation and summarizes the steps described throughout this section for building and deploying the emulator.

\section{Results} \label{sec:results}

\subsection{ROSE performance} \label{subsec:performance}

We present two tests of the performance of ROSE in reproducing the calculations of the traditional high-fidelity solver, both for the individual phase shifts and for the overall differential cross section (see Eqs.~\eqref{Eq:phase shifts} and~\eqref{Eq: diff cs}). The first test over phase-shifts was done to search for possible anomalies~\cite{Lucchese:1989zz}, which are singularities when solutions to Eq.~\eqref{Eq: Scattering Hamiltonian} are calculated using variational principles such as Kohn's~\cite{kohn1948variational} or Newton's~\cite{newton2013scattering}. These anomalies are most apparent when continuously sweeping the parameters or the energy in Eq.~\eqref{Eq: Scattering Hamiltonian}, and have been discussed in several of the recent efforts for constructing reduced order models for nuclear scattering~\cite{furnstahl2020efficient,drischler2021toward,Drischler:2022ipa,melendez2021fast} (see, for example, Fig. 1 of \cite{drischler2021toward}). 

We performed the anomaly search by looking at two different elastic scattering reactions: $^{40}$Ca(n,n) and $^{208}$Pb(n,n), for $\ell \in [0,10]$ in both cases. We choose an emulator configuration of $(n_\phi,n_U)=(14,14)$ wave-function and interaction basis. For each case we defined a high-density energy grid from 5 to 25 MeV -a range that overlaps with anomalies found in previous studies~\cite{drischler2021toward}, and that contains the calibration energy of $14$ MeV we focus on later. Two sets of 50 optical potential parameters (defined in Eq.~\eqref{eqn:params_definition}) were created, a training set and a test set, both obtained by using a Latin Hypercube sampling routine~\cite{mckay1979comparison} centered on the appropriate Koning-Delaroche global parametrization value~\cite{koning2003local}. The width of the sampling boundary was set to 40$\%$ of the central value to increase the covered ground during the exploration. The training set was used to build the emulator at each step of the energy grid and the test set was used to obtain the phase shifts at each energy. These phase shifts were then compared to those obtained by the high-fidelity solver. No evidence of anomalies was found in the search, with the emulator's error in the phase shifts of less than $10^{-3}$ radians.  In Appendix~\ref{App: anomalies} we provide an argument for the absence of observed anomalies in our search based on the condition number of the reduced equation's matrices built using a principal component basis Eq.~\eqref{Eq:POD_basis}, in contrast to a basis made of direct snapshots. Figure~\ref{fig:anomalies} in the same appendix shows the results for the anomaly search for $^{40}$Ca(n,n). 

\begin{figure}[!h]
    \centering{\includegraphics[width=0.5\textwidth]{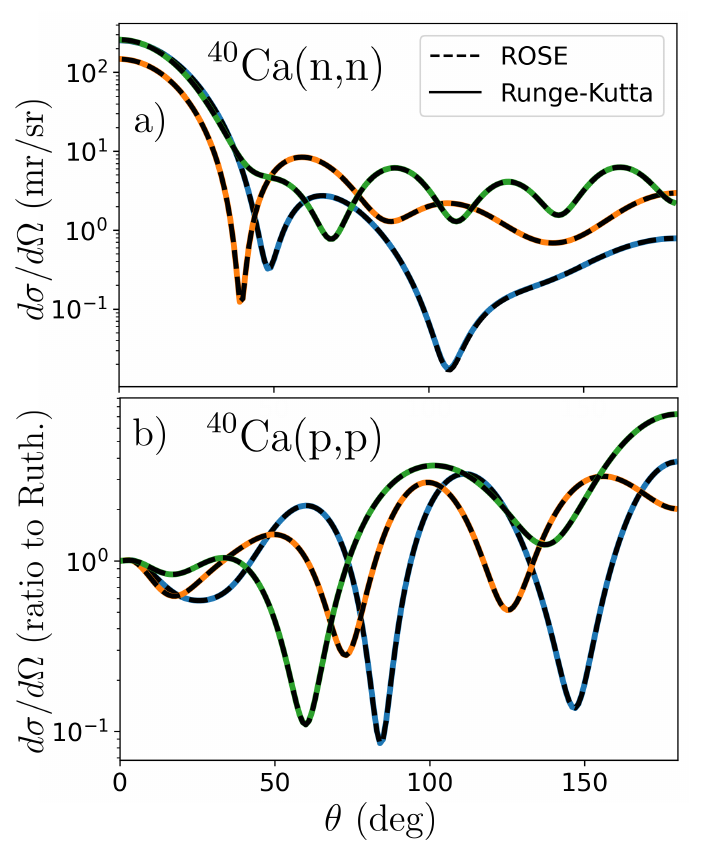}}
    \caption{Calculated differential cross section for $^{40}$Ca(n,n) (a) and $^{40}$Ca(p,p) (b) at 14 MeV for three parameters configurations from the respective test set selected to reflect the range of outcomes. The high-fidelity calculations obtained through Runge-Kutta (colored solid lines) are well reproduced by the emulator with configuration $(n_\phi,n_U)=(15,15)$ (black dashed lines).}
      
    \label{fig:CS_tests}
\end{figure}

\begin{figure*}[!ht]
    \centering{\includegraphics[width=1\textwidth]{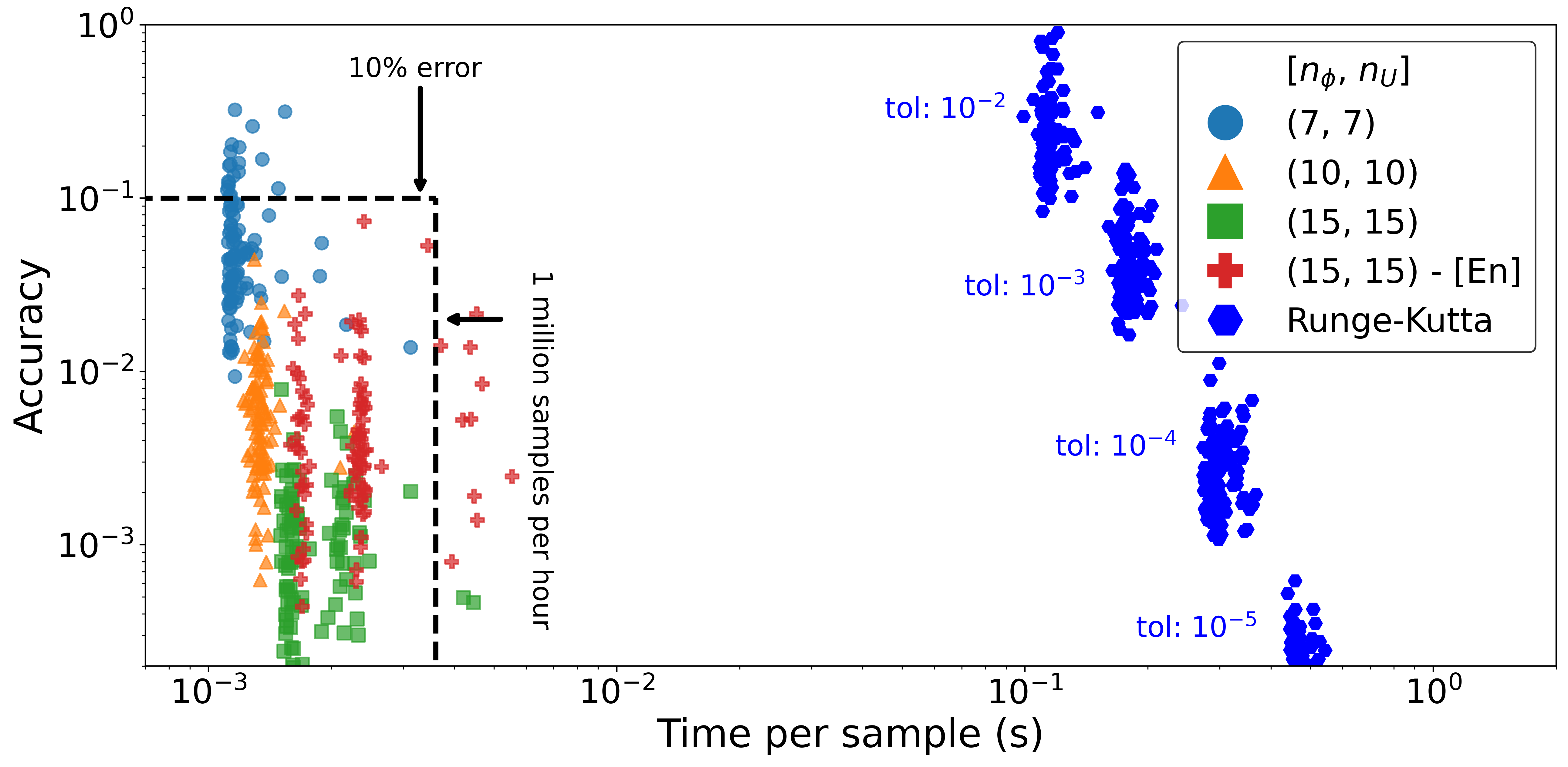}}
\caption{Computational Accuracy vs.\ Time Plot showing the trade-off between accuracy and speed of both ROSE and the high-fidelity solver when calculating the differential cross section of $^{40}$Ca(n,n) at 14 MeV for 100 test parameter values. The $x$-axis represents the time taken for each evaluation, while the $y$-axis represents the Accuracy defined in Eq.~\eqref{Eq: relative error} as the maximum relative error in the entire differential cross section when compared to a high-fidelity solver set to very high precision. For ROSE, we varied the number of basis of the wave functions $n_\phi$ Eq.~\eqref{Eq: RB} and the number of basis in the potential $n_U$ Eq.~\eqref{Eq: potential RB} between 7 and 15. The red crosses, (15,15) - [En], represent an emulator trained across the energy window $[10-30]$ MeV, while the other three configurations were trained at the same 14 MeV energy. For the high-fidelity solver, which uses the Runge-Kutta method, we varied the relative and absolute tolerance between $10^{-2}$ and $10^{-5}$, numbers that are shown next to the blue hexagons representing their performance. The very high precision high-fidelity configuration was $10^{-9}$. The box delimited by dashed lines represents the target zone consisting of, as explained in the text, more than one million samples per hour with less than $10\%$ relative error in the differential cross section. ROSE's performance enters the box and is clearly superior to the traditional high-fidelity method, surpassing its speed by more than two orders of magnitude for a comparable accuracy.}      
    \label{fig:CAT}
\end{figure*}

The second test focuses on the performance of ROSE in reproducing the high-fidelity calculation of the differential cross section---for which all partial waves contribute---for either neutrons or protons impinging on $^{40}$Ca at 14 MeV center-of-mass energy. We chose a maximum partial wave of $\ell=10$ in our calculations, sufficient for convergence. The results of this test guided our selection of the emulator configuration $(n_\phi,n_U)$ for the Bayesian calibration we discuss in the next section. We used two parameter sets, a training and a testing set, each of 200 and 100 parameters, respectively, drawn using the same Latin Hypercube sampling in a box of $30\%$ variations around the prior mean values (see Table~\ref{tab:parameters-calibration neutrons} for neutrons, and Table~\ref{tab:parameters-calibration protons} in Appendix~\ref{App: Results Extras} for protons). Figure~\ref{fig:CS_tests} shows the differential cross sections calculated using the high-fidelity solver (solid colored lines) and ROSE with $(n_\phi,n_U)=(15,15)$ (dashed black lines) for three optical potential parametrizations belonging to the respective test sets. Even though the cross sections for both neutrons and protons vary appreciably, the emulator is able to reproduce the high-fidelity calculations very well.

Figure~\ref{fig:CAT} compares side-by-side the performance of ROSE and the high-fidelity solver running on the same system (commodity hardware). The computational accuracy vs.\ time plot presents the trade-off between precision and speed of both methods when compared against a calculation with the default version of the high-fidelity solver (tolerance of $10^{-9}$). We characterize the accuracy of each method by the maximum relative error over the calculations across every angle between 1 and 179 degrees:
\begin{equation}
    \begin{aligned}\label{Eq: relative error}
        &\text{Accuracy}=\\
        &\text{Max}_{ \{\theta\in [1,179] \} } \Big| \frac{d\sigma/d\Omega_\text{HF}(\theta) - d\sigma/d\Omega_\text{method}(\theta )}{d\sigma/d\Omega_\text{HF}(\theta)}  \Big|,
    \end{aligned}
\end{equation}
where $d\sigma/d\Omega_\text{HF}(\theta)$ refers to the differential cross section calculated with the 
default tolerance ($10^{-9}$), while $d\sigma/d\Omega_\text{method}(\theta )$ refers to the same calculation with either ROSE or a lower tolerance high-fidelity solver.

\begin{figure*}[!htb]
    \centering{\includegraphics[width=1\textwidth]{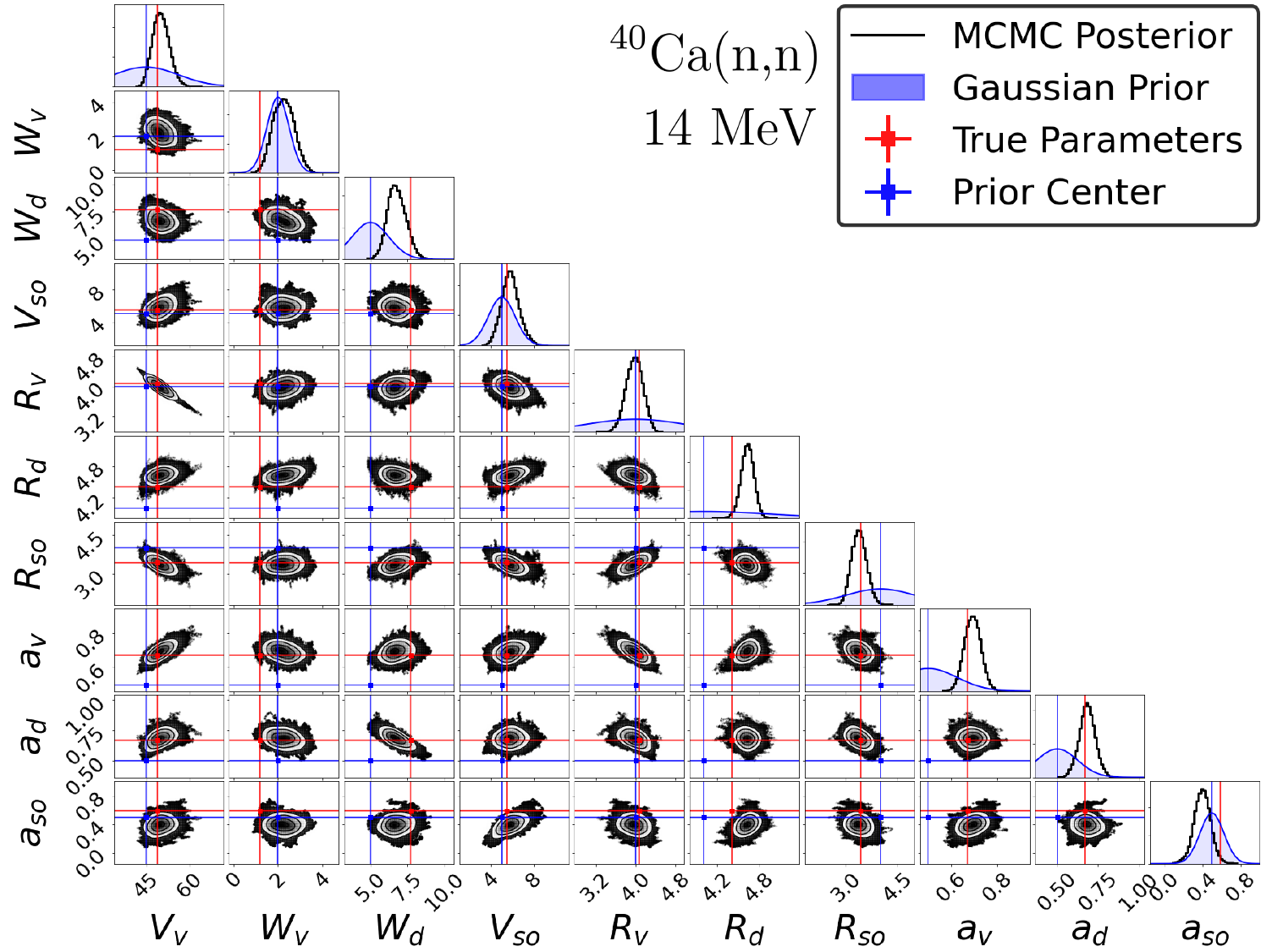}}
    \caption{Cornerplot~\cite{corner} for the calibration of the optical model through the $^{40}$Ca(n,n) reaction at 14 MeV. This plot shows all the one and two dimensional projections of the approximated posterior probability distributions Eq.~\eqref{Eq: posterior emu}. The black histograms represent the posterior Eq.~\eqref{eqn:bayes-posterior}, approximated by 1 million samples visited by 20 MCMC walkers, while the blue filled contours represent the Gaussian prior Eq.~\eqref{eq:prior}. The red lines show the values of the true parameters $\trueparams$ obtained from \cite{capote2009ripl} and used to generate the data while the blue lines show the prior means $\priorparams$. Both sets of parameters are defined in Table~\ref{tab:parameters-calibration neutrons}.}
      
    \label{fig:MCMCNeutrons}
\end{figure*}

As the number of wave-function basis $n_\phi$ and the number of interaction basis $n_U$ increase, the ROSE calculations become exponentially more accurate, diving into the dashed target box of a maximum relative error of $10\%$ (see Sec.~\ref{subsec:calibrationsetup} below), while maintaining a calculation speed of more than 1 million samples per hour, around 1-3 milliseconds per calculation, fast enough for the desired Bayesian calibration. In contrast, as we lower the tolerance of the Python high-fidelity solver (Runge-Kutta), this method does not get near the target box. For comparison, FRESCO~\cite{fresco}, the off-the-shelf traditional solver implemented in Fortran and widely used by reaction practicioners, takes around 30 milliseconds per calculation using standard settings.

The optical potential parameters have energy dependence as a consequence of simplifications in the reaction model.
Thus, energy-dependent total and reaction cross sections provide important constraints on 
these phenomenological potentials, and the ability to train an emulator capable of capturing the resulting energy-dependent behavior of the scattering wavefunctions is desirable for their calibration. The ROSE configuration $(15,15)$ - [En] (red crosses in Fig~\ref{fig:CAT}) represents the performance of an emulator that was trained on the same parameter set as the other cases, but where the energy was varied between 10 and 30 MeV in the training samples. This emulator was then used to calculate the cross section of $^{40}$Ca(n,n) at 14 MeV of the test set, resulting in an accuracy comparable to the single-energy $(n_\phi,n_U)=(10,10)$ version, while maintaining an average calculation speed of around 2--5 milliseconds. These results showcase the ability of ROSE to effectively emulate across energies, an important feature for future energy-dependence calibrations of the optical model.

\subsection{Bayesian calibration} \label{subsec:calibrationsetup}

Having tested ROSE's performance, 
we set up the following calibration task as a demonstration.
Using the high-fidelity solution at a parameter referred as the true parameter $\trueparams$, we collect the true observations as $\obs_0 = \optpot(\trueparams)$.  We generate synthetic data $\obs = \obs_0 + \boldsymbol{\varepsilon},$ where the error $\boldsymbol{\varepsilon}$ are independent Gaussian noises with mean $\boldsymbol{0}$ and standard deviation at $\approx 10\%$ of the observation's value. Following the previous section, the observations in this case are the differential cross section (see Eq.~\eqref{Eq: diff cs}) for $^{40}$Ca(n,n) and $^{40}$Ca(p,p) (separately) at 14 MeV center of mass energy. The 10$\%$ error scale for the elastic differential cross section is a typical magnitude for the experimental error associated with this observable. The final data consists of the differential cross section calculated with the high-fidelity solver at 28 angles between 20° and 155° in spacings of 5°, enough to capture the diffraction pattern.

Based on the performance shown in Fig.~\ref{fig:CAT}, we select the emulator built with $(n_\phi, n_U) = (15, 15)$,
since it offers in the worst case an error of $1\%$ -almost an order of magnitude less than the ``observation'' error of 10$\%$- while still being computationally efficient. To be conservative we set the overall emulator error to $1\%$ across all angles, independent of the value of the parameters. The covariance matrix $\emucov$ is therefore a constant diagonal matrix of size $28 \times28$ with elements $0.01\obs$. We note that careful quantification of this error can be essential---especially if it can increase to a size comparable to the observation error as the parameter exploration diffuses away from the emulator's training region. Nevertheless, for the purpose of this study, we proceed with the conservative constant estimate we described.  

We choose an independent Gaussian prior distribution for each parameter with mean $\singleprior_k$ and standard deviation $0.25 \singleprior_k$, which leads to an overall prior
\begin{equation} \label{eq:prior}
    p(\params) = \prod_{k=1}^{\numparams} p(\singleparam_k),~
    p(\singleparam_k) \propto \exp\left\{-\frac{1}{2} \frac{(\singleparam_k - \singleprior_k)^2}{(0.25 \singleprior_k)^2}\right\}.
\end{equation}
The prior means, as well as the true parameters used for generating the data, are shown in Table~\ref{tab:parameters-calibration neutrons} for neutrons, and in Table~\ref{tab:parameters-calibration protons} in the Appendix for protons.

\begin{figure}[!htb]
    \centering{\includegraphics[width=1\columnwidth]{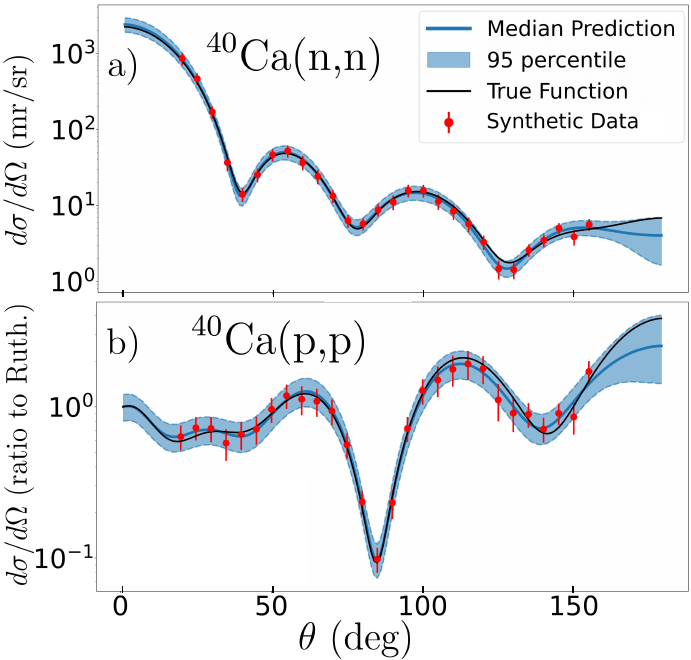}}
    \caption{Predictive posterior distribution for the differential cross section in a) $^{40}$Ca(n,n) 
    and b) $^{40}$Ca(p,p) 
    at 14 MeV. 
    The differential cross sections were calculated using 50,000 random parameters obtained
    during the MCMC sampling in Figs.~\ref{fig:MCMCNeutrons} and~\ref{fig:MCMC Protons}. The $95\%$ credible interval is calculated from Eq.~\eqref{eqn:predictive-posterior} by taking into account the error structure of the data. The synthetic data with a 95$\%$ error bar (in red) was created from Eq.~\eqref{eq: obs} by using the true parameters (black curve) defined in Tables~\ref{tab:parameters-calibration neutrons} and~\ref{tab:parameters-calibration protons}.}
      
    \label{fig:PredictivePostNeutrons}
\end{figure}

\begin{table}[h]
\centering
\begin{ruledtabular}
\begin{tabular}{ldd}
\textbf{Parameter} & \multicolumn{1}{c}{\textbf{Prior mean} $\priorparams$} & \multicolumn{1}{c}{\textbf{True value} $\trueparams$} \\
\midrule
$V_{v0}$ (MeV) & 45 & 48.9 \\
$W_{v0}$ (MeV) & 2 & 1.2 \\
$W_{d0}$ (MeV) & 5 & 7.7 \\
$V_{so}$ (MeV) & 5 & 5.5 \\
$R_{v0}$ (fm) & 4 & 4.07 \\
$R_{d0}$ (fm) & 4 & 4.41 \\
$R_{so}$ (fm) & 4 & 3.42 \\
$a_{v0}$ (fm) & 0.5 & 0.67 \\
$a_{d0}$ (fm) & 0.5 & 0.67 \\
$a_{so}$ (fm) & 0.5 & 0.59 \\
\end{tabular}
\end{ruledtabular}
\caption{The prior means $\priorparams$ and the true values $\trueparams$ for the Optical Potential parameters used to create the data for $^{40}$Ca(n,n) at 14MeV center of mass energy. The exact values were taken from \cite{capote2009ripl}.} 
\label{tab:parameters-calibration neutrons}
\end{table}

The calibration of the optical potential parameters was carried out with \texttt{surmise}, a Python package---part of the BAND framework \cite{bandframework}---that interfaces Bayesian emulation and calibration \cite{surmise2023}.  
The design of \texttt{surmise} facilitates the seamless integration of ROSE as an emulator.  
We employ the Metropolis-Hastings (MH) algorithm available in \texttt{surmise} to sample from the posterior distribution \cite{gelman2013bayesian}.  Details of the integration between ROSE and \texttt{surmise} are given in Appendix~\ref{appx:rose-surmise}, and a summary of the MH algorithm is given in Appendix~\ref{App:metropolis-hastings}.

A total of 20 chains, of 50,000 MCMC samples each, were obtained within an hour on the same commodity computer used to produce Fig.~\ref{fig:CAT}. Each walker had a burn-in period of $n_\text{burn}=3,000$ samples, and started randomly within a small region around the prior center to avoid known multi-modal posteriors of the optical potential~\cite{hodgson1971nuclear}. Figure~\ref{fig:MCMCNeutrons} shows the results of the posterior sampling for neutrons, while Fig.~\ref{fig:MCMC Protons} in Appendix~\ref{App: Results Extras} shows the equivalent results for protons. As has been identified in previous studies, e.g. \cite{lovell2017}, some parameters are strongly correlated - such as the real volume $V_v$ and real radius $R_v$, and they display posteriors that are sharply peaked in comparison to the original prior distribution. Meanwhile, the posterior of other parameters, such as the imaginary volume strength $W_v$, remains close to their prior distribution, not learning much from this specific cross section data at the selected energy. Most of the parameters' true values (red lines) are covered by posterior distribution, with $W_d$, $R_d$, and $a_{so}$ being covered just barely by the tails of the distributions. Further Bayesian studies, powered by emulators such as ROSE, on how much the optical parameters can be constrained by data could be particularly relevant for the new rare isotope beams era \cite{hebborn2023optical}.

The posterior distribution for some of the parameters moved appreciably away from the prior center, the region in which the emulator was originally trained. To verify that our estimate for the emulator error remained applicable we drew 100 parameter values $\params$ from the visited posterior samples and compared the accuracy of the used emulator against the high-fidelity solver. The root mean squared error across the 100 parameters, defined as the square root of the average of the squared residuals between the emulator and the high fidelity, was around 1$\%$ for all the angles $\theta$ for both $^{40}$Ca(n,n) and $^{40}$Ca(p,p). This is consistent with the original valued we assigned $\emucov$, yet it is higher than the typical error displayed in Fig~\ref{fig:CAT}, in which $1\%$ represented a worst case scenario. This highlights the importance of quantifying the emulator's error to give credibility to the conclusions of Bayesian studies, especially after the posterior samples have been obtained.

Finally, in Fig.~\ref{fig:PredictivePostNeutrons} we show the predictive posterior distribution, calculated through Eq.~\eqref{eqn:predictive-posterior}, for $^{40}$Ca(n,n) and $^{40}$Ca(p,p). These cross sections were calculated through 50,000 random samples from the visited parameters on the respective posteriors for both nucleons and a $10\%$ uncorrelated Gaussian noise was added to each value, following the error structure in the likelihood Eq.~\eqref{eqn:bayes-likelihood}. The 95$\%$ credible interval covers very well the data, with all red data points covered and a couple of them almost outside the band, indicating a not overly conservative credibility estimate. Furthermore, the median prediction (solid blue line) overlaps well with the true generating cross section in each case (solid black line).

\section{Conclusions and outlook} \label{sec:conclusion-outlook}

In this paper we presented and showcased a reduced basis emulator workflow for two-body scattering that is applicable to potentials with non-affine parameter dependence, including variations in the beam energy for neutral projectiles. Such non-affine structures, typical in optical reaction models, precludes an immediate offline-online decomposition on emulators based on Galerkin projections or variational principles, potentially limiting their performance. 
The emulator we presented, equipped with the empirical interpolation method to recover an affine structure in the potential parameters, is able to calculate more than a million differential cross sections per hour with a sub-percent relative error using commodity hardware. This performance makes the emulator suitable for Bayesian calibration of the involved reaction models, a very relevant task for the current landscape of direct nuclear reactions~\cite{hebborn2023optical,pruitt2023uncertainty}.
    
A new extensible, user-focused software, \texttt{ROSE}, was developed to construct and implement the emulator. ROSE works together with the Bayesian inference software \texttt{surmise} within the publicly available BAND cyberinfrastructure framework~\cite{bandframework}. This framework is designed for principled uncertainty quantification, including calibration, model mixing, and experimental design,  all of which generally require such emulators. 

In the future, near-term extensions to ROSE for two-body scattering will include allowing for general spin and mass configuration for the projectile and target, the inclusion of non-local potentials and coupled channels, extending the energy emulation for electrically charged projectiles, and a user-friendly functionality for Bayesian calibration of global potentials across energies and isotopes. Developing a framework for better quantification of the emulator error in the online stage, both through statistical studies of the residuals (for example as done in~\cite{melendez2019quantifying}), and through \emph{a posteriori} error estimation~\cite{hesthaven2016certified}, will also become fundamental as the use of ROSE increases. Further exploration of the appearance (or absence) of Kohn anomalies will also be performed, including their possible characterization before the emulator is deployed in terms of the condition number of Eqs.~\eqref{Eq: matrices hat}.

Two-body scattering is the simplest model to describe a reaction when two nuclei collide. With more advanced methods for solving few-body dynamics, each solution becomes much more computationally intensive (e.g.,~\cite{deltuva2007,lei2019,moschini2019,gomez2022}) and generating a large number of high-fidelity samples becomes prohibitive.
A recent application of data-driven emulation (using Gaussian processes) for three-body breakup reactions demonstrates the usefulness of three-body emulators for Bayesian UQ~\cite{surer2022}.
Full three-body emulation is planned for a future release of ROSE.

Finally, we intend to leverage cloud-enabled deployments of ROSE to facilitate engagement across a broad spectrum of interested users and permit online continuous calibration of reaction models in concert with the other tools in the BAND framework.
By deploying trained emulators in cloud environments, one lowers the total computational cost required to evaluate optical models with quantified uncertainties.
This makes for a useful paradigm in educational contexts like training workshops and traditional coursework, but also in lowering the total technological barrier to contribute meaningfully to research in nuclear reactions.
By pairing this with user-focused nuclear science gateways like the Bayesian Mass Explorer web application~\cite{bmex}, both the accessibility of the numerical simulations and the physical results can be enhanced.
A cloud-native implementation of emulation modules also allows for robust interoperability with modern continuous integration pipelines which we intend to integrate with a posterior learning and distribution model enabled by normalizing flows~\cite{yamauchi2023normalizing} to improve the transparency and reproducibility of Bayesian UQ studies.
Given the continued investment and expansion of experimental facilities worldwide, this full integration of advanced cyberinfrastructure methodology into theory workflows is a necessity to effectively capitalize on new data in a timely manner.

\section*{Acknowledgements}

We gratefully acknowledge Yanlai Chen for his insights on the selection of empirical interpolation points, and Amy Lovell and Daniel Phillips for useful discussions. We thank Elizabeth Deliysk for logistic support during the developments of this project.

This work was supported by the National Science Foundation CSSI program under award No.~OAC-2004601 (BAND Collaboration~\cite{bandframework}).
The work of R.J.F. was supported in part by the National Science Foundation Award No.~PHY-2209442.
% and the NSF CSSI program under Award
% No.~OAC-2004601 (BAND Collaboration~\cite{bandframework}).
The work of F.M.N. was in part supported by the U.S. Department of Energy grant DE-SC0021422.
The work of K.B. was supported by the Consortium for Monitoring, Technology, and Verification under Department of Energy National Nuclear Security Administration award number DE-NA0003920.
D.O. and P.G. contributed equally to this work.

\clearpage

\appendix

\section{Scattering details}\label{app:scattering}

Once the numerical solutions $\phi(s)$ of Eq.~\eqref{Eq: Scaled Scattering Hamiltonian} are obtained, the respective phase-shifts $\delta_\ell^j$ are computed by matching the calculated $\phi(s)$ outside of the nuclear potential range to the asymptotic expression in terms of Coulomb functions:
\begin{equation}
    \phi(s)_{s\rightarrow \infty} \propto e^{i\delta_\ell^j}\Big(\text{cos}\delta_\ell^j\  F_\ell(\eta,s)+\text{sin}\delta_\ell^j \ G_\ell(\eta,s) \Big).
\end{equation}
It is important to note that the phase-shifts 
depend on all the variables $\vars$ in Eq.~\eqref{eq: vars}, not only the orbital angular momentum $\ell$ and the total angular momentum $j$. Nevertheless, we explicitly highlight those two quantities in the notation throughout the rest of the manuscript since they play an important role in the calculation of the scattering amplitudes (see Eq.~\eqref{Eq: Amplitudes} below).

Both the regular $F_\ell(\eta,s)$ and irregular $G_\ell(\eta,s)$ Coulomb functions are analytic free solutions to Eq~\eqref{Eq: Scaled Scattering Hamiltonian} when $U(s)=0$. The asymptotic behavior is:
\begin{equation}\label{Eq: asymptotic F}
   F_\ell(\eta,s)_{s\rightarrow \infty} = \text{sin}\big(s-\ell\pi/2+\sigma_\ell(\eta)-\eta\text{ln}(2s)\big),
\end{equation}
with $\sigma_\ell(\eta)=\text{arg }\Gamma(1+\ell+i\eta)$, and $\Gamma$ being a Gamma function. $G_\ell(\eta,s)$ has the same behavior, but with a cos instead of a sin. Near the origin, the asymptotic behavior is~\cite{Thompson2009nuclear}:
\begin{equation}\label{Eq: small s}
    F_\ell(\eta,s)_{s\rightarrow 0} = 
    C_\ell(\eta)s^{\ell+1},
\end{equation}
with the irregular Coulomb function behaving as $G_\ell(\eta,s)\approx s^{-\ell}$. The constants $C_\ell(\eta)$ are calculated recursively:
\begin{equation}
    C_0(\eta) =\sqrt{\frac{2\pi\eta}{e^{2\pi\eta}-1}} \text{ and } C_\ell(\eta) =\frac{\sqrt{\ell^2+\eta^2}}{\ell(2\ell+1)} C_{\ell-1}(\eta).
\end{equation}

The matching equation%
\footnote{It is advantageous to use an explicit numerical representation of $F_\ell(\eta,s)$ and $G_\ell(\eta,s)$ for the matching equation for the phase shift, instead of using the asymptotic forms.
Doing so avoids having to identify a value of $\smatch$ far enough that Eq.~\eqref{Eq: asymptotic F} holds.} 
that can be used to obtain $\delta_\ell^j$ is~\cite{Thompson2009nuclear}:
\begin{equation}\label{Eq:phase shifts}
    S_\ell^{j}\equiv e^{2i\delta_\ell} =\frac{H^-_\ell(\eta,\smatch)-aR_\ell H'^{-}_\ell(\eta,\smatch)}{H^{+}_\ell(\eta,\smatch)-aR_\ell H'^{+}_\ell(\eta,\smatch)},
\end{equation}
where $S_\ell^j$ is the associated S-matrix---a notation we will use for compactness, $H^{\pm}(\eta,s)\equiv G_\ell(\eta,s)\pm iF_\ell(\eta,s)$ are the Coulomb-Hankel functions, $\smatch$ is the value of $s$ at which the numerical solution is matched to the asymptotic ones ($\smatch\leq s_\text{max}$), and $R_\ell$ is related to the logarithmic derivative of the numerical solution for the given $\ell$ at the location $s=\smatch$: 
\begin{equation}
    R_\ell\equiv \frac{1}{\smatch}\frac{\phi(\smatch)}{\phi'(\smatch)}.
\end{equation}
The solutions $\phi$ and the respective phase shifts are calculated for several values of the angular momentum $\ell$ up to a selected $L_\text{max}$, after which higher $\ell$ contributions are neglected. For our case of study of spin 1/2 particles on a spin 0 target the only possible values of $j$ are $\ell+\frac{1}{2}$ and $\ell-\frac{1}{2}$. We then can write the two nuclear scattering amplitudes as~\cite{Thompson2009nuclear,herman2007empire}:
\begin{equation}\label{Eq: Amplitudes}
\begin{aligned}
    A(\theta) =&\ f_C(\theta)+ \frac{1}{2ik}\sum_{\ell=0}^{L_\text{max}}P_\ell(\text{cos}\theta)e^{2i\sigma_\ell(\eta)} \times \\
    & \Big[ (\ell+1)\big(S^{\ell+\frac{1}{2}}_\ell-1 \big) +\ell\big(S^{\ell-\frac{1}{2}}_\ell-1 \big)   \Big], \\
    B(\theta) =& \ \frac{1}{2ik}\sum_{\ell=0}^{L_\text{max}}P^1_\ell(\text{cos}\theta)e^{2i\sigma_\ell(\eta)} \Big[S^{\ell+\frac{1}{2}}_\ell -S^{\ell-\frac{1}{2}}_\ell   \Big],   
\end{aligned}
\end{equation}
where $P_\ell(\text{cos}\theta)$ and $P^1_\ell(\text{cos}\theta)$ in $A(\theta)$ and $B(\theta)$ are Legendre and associated Legendre polynomials, respectively. The Coulomb amplitude is:
\begin{equation}\label{Eq: coulomb}
    f_C(\theta) = -\frac{\eta}{2k\ \text{sin}^2(\theta/2)}\text{exp}\big[ -i\eta\text{ln}(\text{sin}^2(\theta/2)) +2i\sigma_0 \big].
\end{equation}
The differential cross section for unpolarized beams of the target particle is calculated as:
\begin{equation}\label{Eq: diff cs}
    \frac{d\sigma}{d\Omega}(\theta) = |A(\theta)|^2+|B(\theta)|^2.
\end{equation}
If the projectile is a neutron, the Coulomb amplitude \eqref{Eq: coulomb} vanishes and Eq.~\eqref{Eq: diff cs} can be used directly to compare to experimental data. When the projectile is a proton, it is customary to compare the ratio of the nuclear differential cross section in Eq.~\eqref{Eq: diff cs} to the Rutherford cross section (scattering of charged particles):
\begin{equation}
   \sigma_\text{Ruth}=|f_c(\theta)|^2 = \frac{\eta^2}{4k^2\text{sin}^4(\theta/2)} .
\end{equation}

\section{ROSE-\texttt{surmise} integration} \label{appx:rose-surmise}
ROSE is introduced as a new emulator in \texttt{surmise}, callable by the keyword \texttt{nuclear-ROSE}.  Due to 
\texttt{surmise}'s modular design, users may introduce new emulators or new sampling strategies and utilize other available tools in \texttt{surmise} for calibration, by conforming to a simple code framework.  We note that it is also possible to utilize other available sampling strategies from \texttt{surmise} to perform parameter calibration, by a simple change of keyword.

The code details for the use of \texttt{surmise} and the calibration setup can be found in the documentation of ROSE, which includes several tutorials on how to conduct a similar Bayesian study \citep{ROSE2023}.

\begin{figure*}[!htb]
    \centering{\includegraphics[width=1\textwidth]{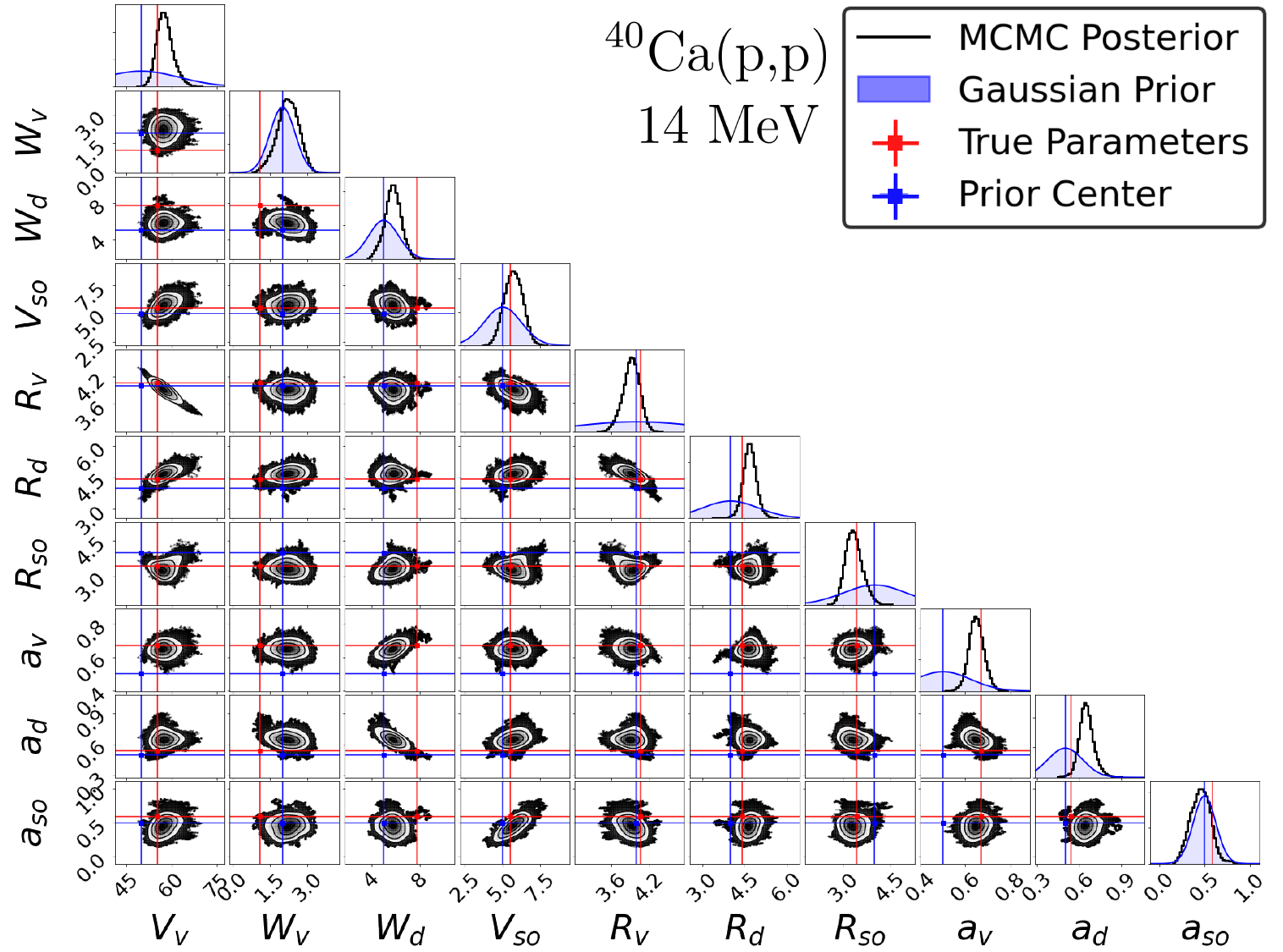}}
    \caption{Cornerplot~\cite{corner} for the calibration of the optical model through the $^{40}$Ca(n,n) reaction at 14 MeV. This plot shows all the one and two dimensional projections of the approximated posterior probability distributions Eq.~\eqref{Eq: posterior emu}. The black histograms represent the posterior Eq.~\eqref{eqn:bayes-posterior}, approximated by 1 million samples visited by 20 MCMC walkers, while the blue filled contours represent the Gaussian prior Eq.~\eqref{eq:prior}. The red lines show the values of the true parameters $\trueparams$ obtained from \cite{capote2009ripl} and used to generate the data while the blue lines show the prior means $\priorparams$.  Both sets of parameters are defined in Table~\ref{tab:parameters-calibration protons}}
      
    \label{fig:MCMC Protons}
\end{figure*}

\section{Metropolis-Hastings algorithm}\label{App:metropolis-hastings}

We detail the Metropolis-Hastings (MH) algorithm in this section. 
The MH algorithm begins
by drawing $\params^{(0)}$ from a shrunken version of the prior distribution Eq.~\eqref{eq:prior} with the same mean but a standard deviation of $5\%$ instead of $25\%$. The optical potential parametrization is multi-modal~\cite{hodgson1971nuclear}, a feature that has to be taken into consideration for uncertainty quantification efforts \cite{hebborn2023optical}. For this study we focus on a single mode by starting the walkers from a small region around the prior. 

\begin{figure*}[!htbp]
    \centering{\includegraphics[width=0.65\textwidth]{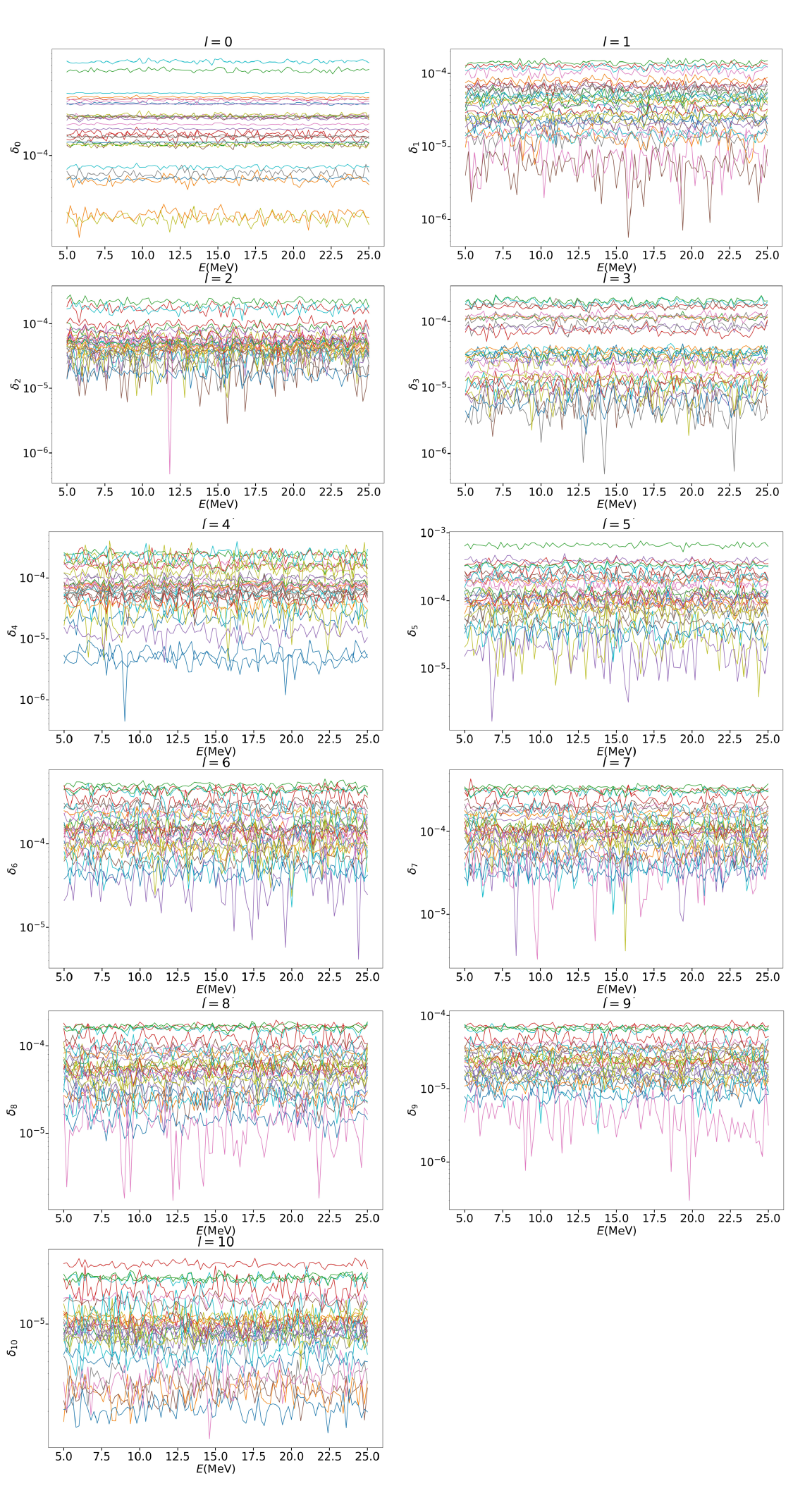}}
    \caption{Residuals in the phase shift when comparing ROSE emulation vs the high fidelity solver across energies for ${40}$Ca(n,n) for $\ell\in[0,10]$. No evidence of Kohn anomalies (see~\cite{drischler2021toward}) was found.}
      
    \label{fig:anomalies}
\end{figure*}

From the current iterate $\params^{(l)}$, the algorithm proposes an iterate, $\params^\prime$, from a multivariate Gaussian proposal distribution that centered at $\params^{(l)}$.  We choose the proposal distribution to be uncorrelated among the parameters, therefore each parameter in the proposed iterate $\singleparam_k^\prime \sim \mathrm{N}(\singleparam_k^{(l)}, \xi_k), k=1,\ldots,\numparams.$ 
The algorithm then decides to accept the proposed parameter as the next iterate, which means that $\params^{(l+1)} = \params^\prime$, with probability dictated by comparing the posteriors Eq.~\eqref{eqn:bayes-posterior}: $\min\{p(\params^\prime|\obs) / p(\params^{(l)}|\obs), 1 \}, $ otherwise $\params^{(l+1)} = \params^{(l)}.$   The values of $\xi$s are tuned to achieve an overall acceptance rate around the range of $0.25$--$0.35$ \cite{roberts2001optimal}. For our study, $\xi_k = 0.007 \singleprior_k$. The MH algorithm generally seeks a higher posterior region and stays around the region via a random walk. The first $n_\text{burn}$ samples of every walker are discarded to ensure that the obtained parameter chains have converged and are representative of the posterior distribution.  The discarding of burn-in samples also reduces the impact of the starting positions on the conclusions of the analysis.

\section{Calibration results for protons}\label{App: Results Extras}

In this appendix we present the calibration results for $^{40}$Ca(p,p) at 14 MeV center of mass energy. These results are analogous to those presented for $^{40}$Ca(n,n) in Sec.~\ref{subsec:calibrationsetup}.

\begin{table}[h]
\centering
\begin{tabular}{lll}
\toprule
\textbf{Parameter} & \multicolumn{1}{c}{\textbf{Prior mean} $\priorparams$} & \multicolumn{1}{c}{\textbf{True value} $\trueparams$} \\
\midrule
$V_{v0}$ (MeV) & 50 & 55.50 \\
$W_{v0}$ (MeV) & 2 & 1.10 \\
$W_{d0}$ (MeV) & 5 & 7.80 \\
$V_{so}$ (MeV) & 5 & 5.50 \\
$R_{v0}$ (fm) & 4 & 4.07 \\
$R_{d0}$ (fm) & 4 & 4.41 \\
$R_{so}$ (fm) & 4 & 3.42 \\
$a_{v0}$ (fm) & 0.6 & 0.67 \\
$a_{d0}$ (fm) & 0.6 & 0.54 \\
$a_{so}$ (fm) & 0.6 & 0.59 \\
\bottomrule
\end{tabular}
\caption{The prior means $\priorparams$ and the true values $\trueparams$ for the Optical Potential parameters used to create the data for $^{40}$Ca(p,p) at 14MeV center of mass energy. The exact values were taken from \cite{capote2009ripl}.}
\label{tab:parameters-calibration protons}
\end{table}

\section{Details on the anomalies study}\label{App: anomalies}

Figure~\ref{fig:anomalies} shows the scan in energies for the anomalies search detailed in Sec.~\ref{subsec:performance} in the main text. One possible reason for the absence of anomalies in our search could be
from the observation that
reduced equations obtained 
by using the orthonormal $\PCAPOD$ basis expansion from Eq.~\eqref{Eq:POD_basis} mitigate ill conditioned matrices, when compared to a basis of snapshots (known as a Lagrange basis~\cite{quarteroni2015reduced}). One way to argue this is to consider the $L^2$ condition number $\kappa(\boldsymbol{M})$ of the matrix $\boldsymbol{M}$ from the reduced equation~\eqref{Eq: matrices}:
\begin{equation}
    \kappa(\boldsymbol{M})=\frac{\lambda_\text{max}(\boldsymbol{M})}{\lambda_\text{min}(\boldsymbol{M})},
\end{equation}
where $\lambda_\text{max}$ and $\lambda_\text{min}$ correspond to the maximum and minimum singular values of the matrix $\boldsymbol{M}$. The condition number $\kappa(\boldsymbol{M})$ encodes information about how many orders of magnitude must be represented to get an accurate solution to the linear system, and it is a metric for the stability of matrix equations \cite{cheney1998numerical}.

We can gain intuition on the condition number of $\boldsymbol{M}$ by writing Eq.~\eqref{Eq: integrals} in matrix notation. Let $\boldsymbol\Phi$ be the matrix with columns equal to the reduced basis elements $\{ \phi_k\}_{k=1}^{n_\phi}$, and let $\mathcal{F}_\alpha$ be the $\mathcal{N}\times\mathcal{N}$ representation of the operator $F_\alpha$, then the matrix  is given by:
\begin{equation}
    \boldsymbol{M}_{[n_\phi\times n_\phi]}=\boldsymbol{\Phi}^\top\mathcal{F}_\alpha\boldsymbol{\Phi}.\label{Eq:Integrals_matrix}
\end{equation}
As shown in \cite{bonilla2022training}, many of the recent variational emulators for scattering are equivalent to constructing $\boldsymbol{M}$ as in Eq.~\eqref{Eq:Integrals_matrix}, but using a Lagrange basis for $\boldsymbol{\Phi}$. In practice, the multicollinearity of the functions used for $\boldsymbol{\Phi}$ leads to an exponential decrease in its smallest singular value (compared to a value of 1 when $\boldsymbol{\Phi}$ is constructed with a $\PCAPOD$ basis). This ultimately decreases the lowest singular value of $\boldsymbol{M}$ through the projection equation~\eqref{Eq:Integrals_matrix} and therefore yield a worse condition number for the system, as we have observed through numerical experiments.

\bibliographystyle{apsrev4-1}

\bibliography{gevc.bib} 
\end{document}